\documentclass[usenatbib]{mn2e}
\usepackage[dvips]{graphicx}
\title[The Non-Parametric Model]{The Non-Parametric Model for Linking Galaxy Luminosity with Halo/Subhalo Mass}
\author[A. Vale and J. P. Ostriker]{A. Vale$^{1}$\thanks{E-mail: avale@ast.cam.ac.uk}
and J. P. Ostriker$^{1,2}$\\
$^{1}$Institute of Astronomy, University of Cambridge, Madingley Road,
Cambridge CB3 0HA, United Kingdom\\
$^{2}$Princeton University Observatory, Princeton University,
Princeton NJ 08544, USA}
\date{Accepted...... Received......; in original form 30 November 2005}

\begin{document}
\maketitle

\begin{abstract}

Non-parametric, empirically based, models for associating galaxy
luminosities with halo/subhalo masses are being developed by several
groups and we present here an updated version of the \citet{paper1}
version of this model.  This is based on a more accurate,
self-consistent treatment of subhalo mass loss and revised results for
the subhalo mass function to address this question anew. We find that
the mass-luminosity relation, at high mass, particularly for first
brightest galaxies and less so for group total, is almost independent
of the actual luminosity function considered, when luminosity is scaled
by the characteristic luminosity $L_*$.  Additionally, the shape
of the total luminosity depends on the slope of the subhalo mass
function. For these high mass, cluster sized haloes we find that total
luminosity scales as $L_{tot}\sim M^{0.88}$, while the luminosity of
the first brightest galaxy has a much weaker dependence on halo mass,
$L_1\sim M^{0.28}$ , in good agreement with observations and previous
results. At low mass, the resulting slope of the mass luminosity
relation depends strongly of the faint end slope of the luminosity
function, and we obtain a steep relation, with approximately $L\sim
M^{4.5}$ for $M\sim 10^{10} h^{-1} M_\odot$ in the K-band. The average
number of galaxies per halo/cluster is also in very good agreement
with observations, scaling as $\sim M^{0.9}$.

In general, we obtain a good agreement with several independent sets
of observational data. Taking the model as essentially correct, we
consider two additional possible sources for remaining discrepancies:
problems with the underlying cosmology and with the observational mass
determination. We find that, when comparing with observations and for
a flat cosmology, the model tends to prefer lower values for
$\Omega_m$ and $\sigma_8$. Within the WMAP+SDSS concordance plane of
\citet{tegmark}, we find best agreement around $\Omega_m=0.25$ and
$\sigma_8=0.8$; this is also in very good agreement with the results
of the CMB+2dF study of \citet{sanchez}. We also check on possible
corrections for observed mass based on a comparison of the equivalent
number of haloes/clusters.  Additionally, we include further checks on
the model results based on the mass to light ratio, the occupation
number, the group luminosity function and the multiplicity function.

\end{abstract}


\section{Introduction}

One of the outstanding challenges of cosmology is to relate the more
theoretical aspect of the standard cosmological model, in the form of
the large scale distribution of dark matter as seen in high resolution
N-body simulations, to the observational evidence, as reflected by
large galaxy surveys. Or, in a directly related question, what is the relation
between observable properties of galaxies and computable properties of 
dark matter haloes?

The traditional route to looking at this problem has been to follow
the theory of galaxy formation. This can be tested through the results
of hydrodynamical simulations
\citep{whs,ytj,pjf,nfco,Berlindetal,meza,bailin}, which combine dark
matter with baryons, or semi-analytical models of galaxy formation
\citep{kns,gbf,kcda,kcdb,bba,bbb,sd,sls,wsb,bfb,Berlindetal}. This is
a powerful way of looking at the problem, since it provides a direct
answer, and a galaxy formation theory is an end goal in itself. There
are, however, some problems: first and foremost, the theory behind
galaxy formation has several components which are ill understood, and
where the best approach remains phenomenological. At the same time,
the required very high resolution, large scale, full hydrodynamical
simulations including magnetic fields and radiative transfer are
beyond current computing resources.

In the past few years, a new approach has appeared which is in many
ways an alternative: this consists of an indirect method, associating
galaxy and dark matter halo properties in an empirical manner. This is
usually done through a statistical approach, either by focusing
directly on the number of galaxies in each halo, as is done in halo
occupation distribution models
\citep{seljak,benson,bws,zt,bg,Berlindetal,mp}, or through the
luminosity distribution of the galaxies in a halo, like the
conditional luminosity function approach
\citep{BoschYangMo,YangMoBosch,vymn,coorayb,coorayc,cooraye}, or
finally by building a direct relation between mass and luminosity
\citep{PeacockSmith,kravtsov,paper1,tasitsiomi}.

In the present paper, we take the latter approach. We use new, high
resolution simulation results for the subhalo mass distribution,
together with a self consistent approximation to subhalo mass loss, to
build the total distribution of dark matter hosts based on physical
theory.  Together with the empirically determined galaxy luminosity
function, this yields a mass luminosity relation for an individual
galaxy and the halo or subhalo which hosts it. This is in part an
updated version of the preliminary model shown in 
\citet[hereafter paper I]{paper1}.

This paper is organised as follows: in section 2, we introduce the
non-parametric model and the main concepts behind it. In section 3, we
look in some detail at the subhalo mass function, introducing a
prescription for subhalo mass loss and in particular looking at how to
define this mass function in terms of the original, pre-merger into
parent, mass of the subhaloes. In section 4, we give the main results
for our base model, using the K-band luminosity function from the
2MASS survey. In section 5, we look at how the model results change
depending on the luminosity function (and hence waveband) used, as
well as the effect of changing the underlying cosmological model and
what we find is the most relevant parameter of the subhalo mass
function, the low mass slope. We also look at potential problems with
the observational mass determination.  We present additional checks of
the model in section 6, by calculating the predicted mass to light
ratio, occupation number, group luminosity function and multiplicity
function. Finally, we conclude in section 7.

\section{The Non-parametric Model}

The basic idea behind the non-parametric model is to assume a
monotonic and one to one relation between the luminosity of a galaxy
and the mass of the dark matter halo or subhalo which hosts it. This
comes from the standard picture that galaxies are formed in haloes
through the accretion of gas, the amount of which is a monotonic
function of the depth of the potential well of the halo and thus of
its mass. By including the subhaloes, we are assuming that all
galaxies are either hosted individually in a parent halo, or in the
case of multiple systems like clusters in one of the subhaloes. In
fact, we assume that counting haloes and subhaloes accounts for all
hosts, and exclude complications such as conditions during ongoing
mergers.  Groups and clusters are then formed when haloes merge; the
end result of such a build up is to have a central galaxy, which
formed in the most massive of the initial haloes (which subsequently
became the parent), and satellite galaxies in the smallest haloes
which were accreted and which became subhaloes.

Using this basic concept, it becomes possible to obtain an average
relation between the luminosity of a galaxy and its host halo/subhalo
mass by matching the numbers of each, the former from large scale
galaxy surveys, the latter from dark matter simulations. The average
luminosity $L$ of a galaxy in a halo or subhalo of mass $M$ will then
simply be given by:

\begin{equation} \label{mldef}
\int_L^\infty \phi(L) dL = \int_M^\infty n(M) dM \, ,
\end{equation}

\noindent where $\phi(L)$ is the galaxy luminosity function and
$n(M)=n_H(M)+n_{SH}(M)$ is the sum of the halo and subhalo mass
functions.

During and after the halo merging process significant star formation
in the subhalo declines (e.g., \citealt{kwg,sp,cole}), so we will take
as a constraint that the amount of gas accreted by what becomes a
subhalo, and which will thus be available to form a satellite galaxy,
will be proportional not to the mass it has at present, but to the
maximum mass it had before being accreted by its parent halo and
undergoing mass loss to tidal stripping and dynamical friction.  This
is supported by some recent work of \citet{libeskind}, who use high
resolution N-body simulations together with semi-analytical modelling
of the formation of galaxies to study the distribution of satellites
in Milky Way type galaxies. They find that while the spatial
distribution of satellite galaxies is significantly different from
that of the most massive present day subhaloes, it is well matched by
that of the subset of subhaloes with the most massive progenitors. It
is then necessary to use not the distribution of subhaloes as a
function of their present mass, but instead of the original mass they
had prior to accretion into the parent and subsequent mass loss,
regardless of theoretical issues related to galaxy formation. We
circumvent this problem by coupling the present, evolved subhalo mass
function with a prescription for the amount of mass loss, together
with some simple arguments on the total mass contained in these
subhalo progenitors, in order to regain the initial mass of a subhalo
at the moment of accretion and tag each of these by its initial mass.

\section{Subhaloes}

\subsection{Mass loss}

As discussed, the first step to having a workable non-parametric model
is to account for the amount of mass loss in subhaloes. To obtain it,
we adapt the results of \citet{vdbshmf}, by comparing the results they
give for the evolved and unevolved (original) subhalo mass
functions. If we assume that the present and original masses of the
subhaloes are related monotonically (that is, that the most massive of
the original subhaloes are still the most massive at present,
independently of the actual amount of mass loss), we can obtain this
relation by comparing the total numbers of present subhaloes to their
original progenitors, in a manner similar to the process presented
above for the non-parametric model. 

\begin{figure}
\includegraphics[height=84mm,angle=270]{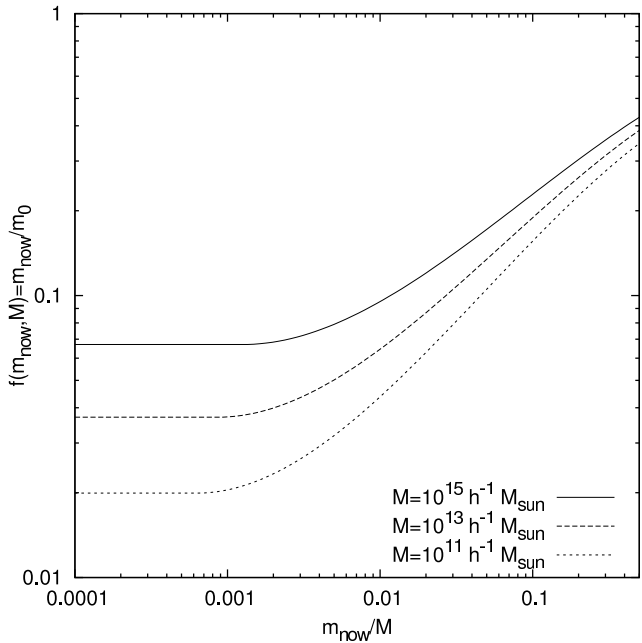}
\caption{Mass loss factor $f(m_{now}/m_0)$, which is a function both of
subhalo and parent halo mass. Plotted are curves for three different
parent halo masses, $10^{15} h^{-1} {\rm M_\odot}$, $10^{13} h^{-1}
{\rm M_\odot}$, $10^{11} h^{-1} {\rm M_\odot}$, from top to bottom
respectively. These curves have the normalization we obtained directly
from adopting the results of \citet{vdbshmf}; it may be necessary to
renormalize them when considering different present subhalo mass
functions (see text for discussion).  }
\label{massloss}
\end{figure}

The resulting mass loss factors will depend on both the subhalo mass
and the mass of the parent halo, as can be seen in figure
\ref{massloss}. In fact, we have made a small further change to the results
we obtain from the mass functions of \citet{vdbshmf}, which is to
flatten the mass loss factor to a constant at its minimum and so maintain
monotonicity. The
upturn at high subhalo mass is explained by two competing factors
which contribute to mass loss: while the more massive subhaloes lose a
larger fraction of their mass in each orbit within the parent halo,
they will also, on average, have formed and therefore been accreted
later, which means they will have undergone fewer of these orbits than
less massive subhaloes (see for example, \citealt{vdbshmf}). 

We can now apply this mass loss factor to subhalo mass functions
measured from simulation results to obtain the original subhalo mass
function. There is an important further consideration, though: care
must be taken with the normalization of this function. In fact, we
know that the sum of the mass in these original subhaloes must equal
the total present mass in the parent halo, since the latter is built
up by accreting and stripping subhaloes (ignoring for simplicity the small
amount of mass associated with subhaloes that disappear totally). Therefore, when using
different subhalo mass functions, the result presented in figure
\ref{massloss} should be taken only to be the shape of the mass loss
factor.  It is then necessary to either renormalize the mass loss
factor (or alternatively the derived original subhalo mass function)
to ensure that the total mass in these original subhaloes matches the
present mass of the parent halo.

\subsection{The subhalo mass function}

The distribution of the subhaloes is based on the subhalo mass
function (SHMF), $N(m|M)dm$, which gives the number of subhaloes in
the mass range $m$ to $m+dm$ for a parent halo of mass $M$. We start
by defining the present day mass fraction in subhaloes as:

\begin{equation} \label{fractiondef}
\gamma(M)\equiv\frac{1}{M}\int_0^{m_{cut}} m N(m|M) dm \, ,
\end{equation} 

\noindent where $m_{cut}(M)$ is the mass of the most massive subhalo 
possible, which we take to be a function of the parent halo
mass. This value will be set by assuming that the maximum original
mass is half the total mass of the parent, $M/2$. This is in fact a
question of definition, since a subhalo which had a mass greater than
this would actually be larger than any other it could be merging with,
and would itself in fact be the parent halo. This can be converted
back to $m_{cut}$ by using the mass loss factor.

This mass fraction in subhaloes is generally found to be a growing
function of parent halo mass, both in simulations or semi-analytical
models (e.g., \citealt{gao,vdbshmf,laurie}).  As pointed out in
\citet{vdbshmf}, this is naturally due to the later formation of more
massive haloes, therefore allowing less time for mass loss to occur.
There also seems to be some agreement that the value of this mass
fraction should be just under 10\%, even at high halo mass.  This will have
as an important consequence that the subhalo mass function cannot
possibly be universal but must depend on $M$, since the subhalo
mass fraction in effect sets its normalization.

Following what was found by \citet{jochen,laurie}, we can then take the SHMF to
have the form of a Schechter function:

\begin{equation} \label{shmf}
N(m|M) dm = A_M \Big(\frac{m}{\beta
M}\Big)^{-\alpha} {\rm exp}\Big(-\frac{m}{\beta M}\Big) \frac{dm}{\beta M} \, ,\\
\end{equation}
\begin{equation}
\nonumber
A_M=\frac{\gamma(M)}{\beta (\Gamma[2-\alpha]-
\Gamma[2-\alpha,m_{cut}(M)/\beta M])}  \, ,
\end{equation}

\noindent where $\alpha$ gives the low mass slope and $\beta$
represents an additional cutoff mass. Typical values for the low mass
slope $\alpha$ are around $\sim 1.9$ \citep{delucia,gao,vdbshmf,zentner}, 
although \citet{laurie} have found a less steep $\alpha=1.75$ when
fitting a power law and an even flatter $\alpha=1.5$ when fitting to a
Schechter function of the form of equation (\ref{shmf}). We should
also point out that \citet{vdbshmf} obtain a slope which varies
slightly with parent halo mass. The cutoff value has alternatively
been determined as $\beta=0.13$ from semi-analytical models
\citep{vdbshmf}, or $\beta=0.3$ from simulations \citep{laurie}. To
this expression should be added the cutoff we are implicitly considering at
$m_{cut}$, as expressed in equation (\ref{fractiondef}). The first
term on the right hand side guarantees that the mass fraction in
subhaloes is $\gamma(M)$. From this we see that $m_{cut}$ has
essentially an effect on the normalization of the SHMF.

However, as discussed above, what we really require to build the
non-parametric model is the original subhalo mass function instead. To do
this, we apply the mass loss factor presented in the previous section
to the SHMF of equation (\ref{shmf}), that is, we use
$m_{now}=f(m_{now},M) m_0$, with $f(m_{now},M)$ the mass loss
function, renormalizing as appropriate to guarantee that the original
total mass in subhaloes equals the present mass of the parent
halo. In terms of the original mass $m_0$, the SHMF of equation \ref{shmf} then
becomes

\begin{figure}
\includegraphics[height=84mm,angle=270]{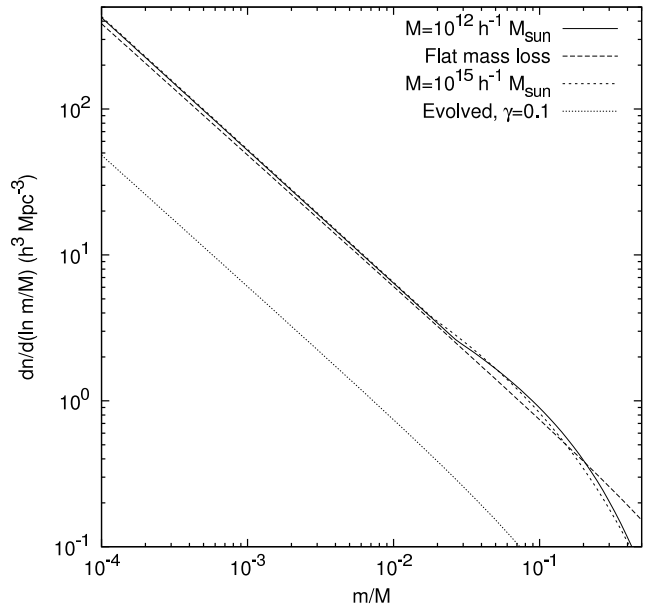}
\caption{ The subhalo mass function (SHMF): shown are the original
  SHMFs for parent haloes of mass $10^{12} h^{-1} {\rm M_\odot}$ and
  $10^{15} h^{-1} {\rm M_\odot}$. For comparison, we also show the
  original SHMF for a $10^{12} h^{-1} {\rm M_\odot}$ halo when we take
  a flat mass loss factor, instead of one based on the curves of
  figure \ref{massloss}, and a present, evolved SHMF with a subhalo
  mass fraction of $\gamma=0.1$.}
\label{shmffig}
\end{figure}

{\setlength\arraycolsep{2pt}
\begin{eqnarray} \label{origshmf}
N_0(m_0|M)dm_0 &=& N[f(m,M)m_0|M]f(m,M)\times \nonumber \\
 & & {} \times \Big(1-\frac{d{\rm log}f(m,M)}{d{\rm log}m}\Big)dm_0 \, ,
\end{eqnarray}}

\noindent where $f(m,M)$ is the mass loss factor, with $m$ the present
subhalo (i.e., the function plotted in figure \ref{massloss}); we have
also explicitly included the transform of the differential term $dm$.
Once more, we are implicitly assuming a cutoff at an original subhalo
mass of $m=M/2$. Figure \ref{shmffig} shows different SHMFs calculated
using this scheme.

There are some important conclusions to be drawn from figure
\ref{shmffig}. First, unlike the case with the present SHMFs, the
original ones are almost universal when plotted as a function of
$m/M$. This is a result of the fact that we are putting in what is, in
these units, a universal normalization at the total halo mass $M$ and
also a universal cutoff at $m/M=0.5$, which usually dominates over the
one in equation (\ref{shmf}), after transforming to the original
mass. As a direct consequence of this, we can see that the original
SHMF, as calculated through this method, is essentially independent of
the present subhalo mass fraction $\gamma$ and largely independent of
the cutoff given by the parameter $\beta$. It is also largely
insensitive to the shape of the mass loss factor as determined in the
previous section.  However, it should be cautioned that this depends
on the actual mass loss factor $f$ used: this is only true if it is
fairly regular, or more precisely $d{\rm log}f/d{\rm log}m\ll 1$, as
is mostly the case of the function we are using presently, which
mostly causes a rescalling of the subhalo mass function, without much
altering its shape.  In such a case, in fact, the only parameter of
the present SHMF which has a large effect is the low mass slope
$\alpha$. This indicates that it should be mostly correct, in the
context of building a non-parametric model like we are doing here, to
use as the original SHMF a power law of slope $\alpha$, with a
cutoff at $m/M=0.5$ and the whole normalized so that the total
original mass in subhaloes equals the present parent halo mass. From
this, it is important to retain that, accepting these two conventions we
are using on the normalization and cutoff, the only element of the
SHMF to which our model is sensitive is the low mass slope.

For the subsequent calculations, we will use a model with the mass
loss factor calculated in the previous section, a present SHMF given
by equation (\ref{shmf}) with $\alpha=1.9$ and $\beta=0.3$. As mentioned
above, the actual value of the mass fraction $\gamma$ is not important
since we have to renormalize the original mass function we obtain; it
is only factored if we wish to have the appropriately normalized mass
loss factor $f(m_{now},M)$ to convert between original and present
subhalo mass using $m_{now}=f(m_{now},M) m_0$. Hereinafter, unless
otherwise stated, all subhalo masses refer to the original mass.

\section{Basic model}

\begin{figure}
\includegraphics[width=84mm]{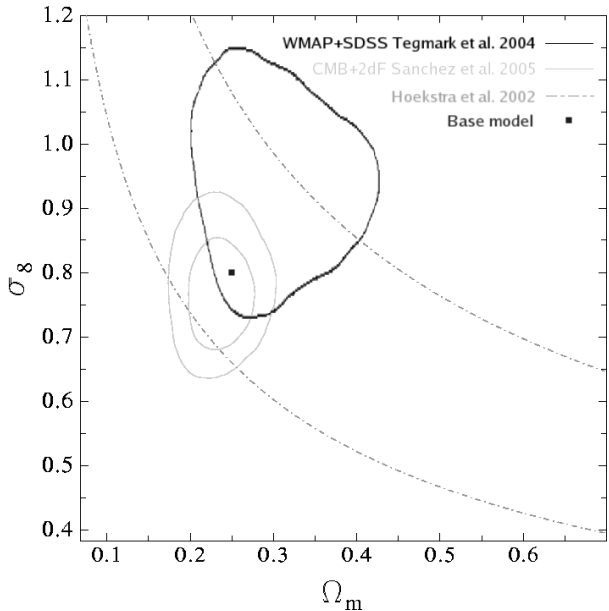}
\caption{ 
The $\Omega_m-\sigma_8$ plane. Shown are the 1- and $2\sigma$ countours from the
CMB+2dF power spectrum results of \citet{sanchez}, together with the $2\sigma$ 
concordance region from the WMAP+SDSS results of \citet{tegmark}. Also shown
as an example of weak lensing results are the constraints of \citet{lensing}.
The point marks the parameter values chosen for our basic model.} 
\label{cosm}
\end{figure}

In this section we present our base model. We take a flat cosmology
 with parameters $\Omega_M=0.25$, $\sigma_8=0.8$ and $h=0.7$. While
 these do not exactly correspond to the current standard model
 \citep{triangle,wmap,tegmark}, they are within the allowed range of
 the $\Omega_m-\sigma_8$ plane determined from the joint WMAP-SDSS
 study of \citet{tegmark}, and we find they produce results better
 matching observations (see section \ref{seccosm} below) than do
 models with slightly higher $(\Omega_m,\sigma_8)$.  On the other
 hand, they match very well with the results found from an analysis of
 CMB and 2dF power spectrum by \citet{sanchez}, being near the center
 of their $\Omega_m-\sigma_8$ concordance region (see figure \ref{cosm}).  
For the basic
 model, we use the luminosity function in the K-band as determined
 from the Two Micron All-Sky Survey (2MASS; \citealt{2massint}). Using
 the K-band allows us to avoid possible complications and
 discrepancies arising from brief intervals of active star formation,
 and therefore helps present a clearer picture. We use here a single
 waveband for clarity in the construction of the model; a comparison
 of results for different wavebands using different luminosity
 functions is presented in section \ref{seclf}.

\subsection{Host mass function}

The first step to building the non-parametric model is to combine the
above subhalo mass function with the parent halo distribution in the
form of the halo mass function to obtain the global distribution of
the dark matter hosts of galaxies. This is also the step where the
cosmological parameter dependency is factored in, by determining the form of
the mass function. We will use the Sheth-Tormen mass function
\citep{stmf}, given by:

\begin{equation} \label{stmf}
n_h(M) dM = A \Big( 1+\frac{1}{\nu^{2q}}\Big) \sqrt{\frac{2}{\pi}} 
\frac{\rho_m}{M} \frac{d\nu}{dM} {\rm exp}\Big(-\frac{\nu^2}{2}\Big) dM\, ,
\end{equation}

\noindent with $\nu=\sqrt{a}\frac{\delta_c}{D(z) \sigma(M)}$,
$a=0.707$, $A\approx 0.322$ and $q=0.3$; as usual, $\sigma(M)$ is the
variance on the mass scale $M$, $D(z)$ is the growth factor, and
$\delta_c$ is the linear threshold for spherical collapse, which in
the case of a flat universe is $\delta_c=1.686$. This mass function
can be roughly approximated by a power law at low mass, scaling as
$\sim M^{-1.95}$, and an exponential cutoff at high mass of the form
$exp(-M/M_*)$, where the cutoff mass is defined by $M/M_*\equiv
\nu^2/2$ and we have roughly $M_*\approx 5\times 10^{14} h^{-1} {\rm M_\odot}$.
The total distribution of dark matter hosts is then given by combining
this with the global distribution of subhaloes, which can be
calculated from the SHMF (\ref{shmf}) and the halo mass function
(\ref{stmf}) by

\begin{equation} \label{globalshmf}
n_{sh}=\int_0^\infty n_h(M) N(m|M) dM \, .
\end{equation} 

\begin{figure}
\includegraphics[height=84mm,angle=270]{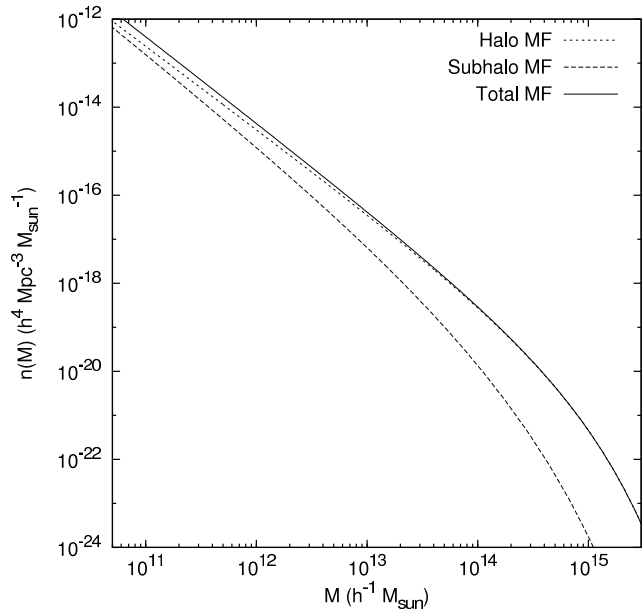}
\caption{Global mass functions for haloes and subhaloes. 
  In the case of the subhaloes, the
  x-axis refers to the original, unstripped mass (see discussion in
  text).} 
\label{mffig}
\end{figure}

\noindent These mass functions are shown in figure \ref{mffig}. Note
that in the combined distribution of haloes/subhaloes the former
dominate with subhaloes approaching parity only on mass scales below
$10^{12} h^{-1} {\rm M_\odot}$.

\subsection{Luminosity Function}

The galaxy distribution is accounted for by the galaxy luminosity
function, which we assume takes the shape of a Schechter function
\citep{schechter76}:

\begin{equation} \label{schechter}
\phi(L) dL = \phi_* \Big(\frac{L}{L_*}\Big)^{\alpha} {\rm
exp}\Big(-\frac{L}{L_*}\Big) \frac{dL}{L_*} \, .
\end{equation}

\noindent For this base model, we use the K-band results from the
2MASS survey, with parameters given by $\alpha=-1.09$,
$\phi_*=1.16\times10^{-2} h^{3} {\rm Mpc^{-3}}$ and $M_*-5 {\rm log}
h=-23.39$ \citep{2mass} (see also table 1).  In section \ref{seclf}
below, we analyse the result of using luminosity functions in other
wavebands. This fit applies to the magnitude range $-22>M_K-5 {\rm
log}_{10}h>-25$.  An important point to note is that we will assume
that the fit actually holds outside of this limited range, and
extrapolate its result outside of it as necessary. This raises the
question of how good a description of the real luminosity function a
Schechter function is and, in particular, whether a power law is a
good fit to the faint end.  Observations at the faint end are plagued
by a host of natural difficulties, most of all the need to sample
efficiently low luminosity and low surface brightness galaxies. It is
not yet entirely certain whether the luminosity function at the faint
end can be well fit by the power law in the Schechter function over a
wide range (see for example \citealt{trentham1,trentham2}; see also
\citealt{blanton}, who look at faint galaxies in SDSS and find that a
Schechter function is a poor fit, and need to introduce a double rather
than single power law parametrization at the faint end). For now, we
will simply assume that we can extend the measured faint-end slope of
the 2MASS survey to fainter magnitudes than their stated limit.

\subsection{Mass Luminosity relation}

\begin{figure}
\includegraphics[height=84mm,angle=270]{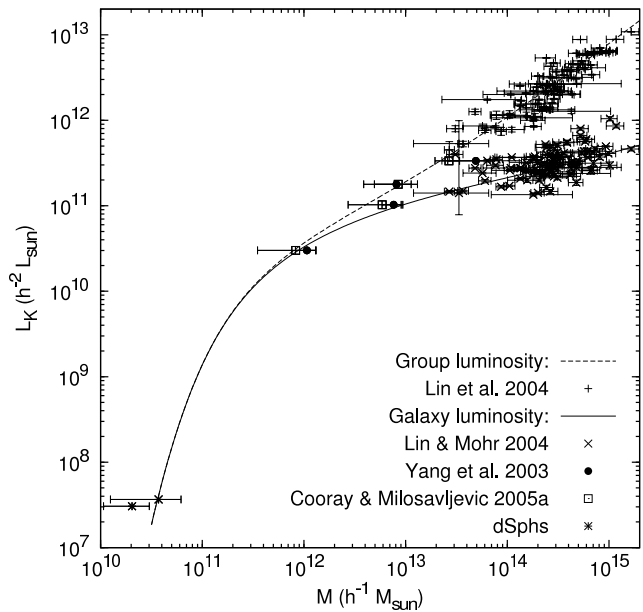}
\caption{Mass K-band luminosity relation, both for single galaxies and
total group luminosity (upper curve). Galaxy luminosity applies to
both haloes and subhaloes, where the mass of the latter refers to the
original, prior to accretion into the parent case. The data points
shown come from \citet{lm} and \citet{lms} (on cluster scales, for
both total and brightest galaxy luminosity), analysis of SDSS weak
lensing \citep{yangwl,cooray}, and an estimate of Milky Way dwarf
spheroidal mass and luminosity (see text for details).}
\label{mlfig}
\end{figure}

We can now combine the total dark matter host distribution with the
galaxy distribution to get a relation between halo/subhalo mass and
galaxy luminosity, by matching counts as described above in section 2. 
The resulting relation is shown in figure
\ref{mlfig}.

Also shown are observational data for brightest cluster galaxy
luminosity and cluster luminosity versus cluster mass, taken from the
K-band results of \citet{lm} and \citet{lms}. These studies use X-ray
temperature results to determine the cluster mass together with
luminosities taken from the 2MASS survey; we include a correction to
the halo mass to account for a virial overdensity of 100 times the
critical density (e.g., \citealt{bn}), instead of 200, based on taking
an NFW halo with a concentration of 5 as was done in \citet{lms}
(see further discussion in section \ref{secmobs}). The points at the
low-end come from Milky Way dwarf spheroidal satellites, where the
mass was estimated from results of \citet{Hayashietal} (see also
discussion in paper I) and the luminosity comes from the
B-band results of \citet{Mateo}, then using an average $B-K=3.6$ to
convert to the K-band \citep{mt}. The intermediate mass results come
from an analysis of SDSS weak lensing results in \citet{yangwl} and
\citet{coorayb}; in both of these, we have transformed the z-band
luminosity to K-band by using ${\rm log}(L_K/L_z)=-0.014+0.492(g-r)$, with
an average $(g-r)=0.6$ \citep{bell}. For all these cases, we again
converted to $M_{100}$, using for the concentration the
\citet{bullock} model.

We find that the mass luminosity relation we obtain (figure \ref{mlfig}) 
can be well approximated by a double power law of the type:

\begin{equation} \label{fit}
L=L_0\frac{(m/m')^{a}}{(1+(m/m')^{b k})^{1/k}} h^{-2}
{\rm L_\odot}\, ,
\end{equation}

\noindent with the parameters $L_0=1.23\times10^{10}$,
$m'=3.7\times10^{9} h^{-1} {\rm M_\odot}$, $a=29.78$, $b=29.5$, and
$k=0.0255$. This fit is good to within $\sim 6\%$ in the range
$5\times 10^{10}<M/(h^{-1} {\rm M_\odot})<5\times10^{15}$. The most
important fact to retain from this fit is that luminosity scales as
$L\propto M^{0.28}$ at high mass. The high values obtained for the
exponents $a$ and $b$ are an artifact of the fact that the relation is
steepening as the mass decreases (note the value of the break mass
given by $m'$).  Group luminosity scales as $L_{group}\propto
M^{0.88}$ for high halo mass.  Overall, these parameters look
different from the fit to observed data done using the same functional
form as equation (\ref{fit}) by \citet{coorayb}.  The main reason for
the difference is the corrections we put in for the mass
estimates. Also, as noted, care should be taken with the low mass
slope, since the values for the best fit curve depend on the lowest
mass used for the fit.  It is, however, easy to predict this slope: if
the mass function goes as $\sim M^{-a} dM$ (where this represents the
total for haloes plus subhaloes) and the luminosity function
as $\sim L^{-b} dL$, then the luminosity will scale as $L\propto
M^c$, with $c=(a-1)/(b-1)$. With $a\simeq1.95$ and $b=1.09$, 
we get $c\simeq10.5$.

Overall, these results seem a good match to the observations. 
The quality of this match has improved over the results we presented in
paper I, especially for the group/cluster luminosity. As we had
hinted in the discussion in paper I, this is most likely due to the
fact that we now have a better, more self consistent, way of treating
the subhalo mass fraction and mass loss. It is also interesting to
note that, since we obtain a total luminosity which scales almost linearly 
with mass, we naturally obtain a mass to light ratio on cluster scales
that is almost constant; this is intrinsic to the model, and is in
good agreement with observations (see section \ref{secmlratio}
below). It is also worth to point out that the agreement for the group
luminosity is not trivial: only the halo/subhalo mass-galaxy
luminosity relation is constructed directly from the model. The group
luminosity is obtained by assuming that applying this derived galaxy
luminosity to the system of the haloes and their subhaloes will result
in a good match for the corresponding galaxy systems. The agreement we
obtain seems to point out that this is indeed a valid assumption (but see
\citealt{cooraycen} for some potential problems involving the detailed
luminosity functions for individual groups when following this
prescription). 

There is some discrepancy between our results and the results
for the higher mass bins based on weak lensing of SDSS galaxies (the points
from \citealt{yangwl,cooray}). While there is no straightforward way to 
explain the differences we find, we caution that there are several factors
that, when taken together, may be enough to account for them. Most relevant
of these are the fact that we are using an average colour to change the 
luminosities from the z- to the K-band, and the possibility that the 
mass being accounted for does not correspond exactly to the definition
we are using. Intriguingly, these weak lensing results seem to be a good 
match to our curve for the total luminosity, which raises the additional 
possibility that the observed luminosity includes not only the central
lensing galaxy, but also some close, faint, satellites (since we would expect 
that most substructure in these medium mass parents to be relatively small).

\section{Dependence on LF and cosmology}

\subsection{Luminosity function} \label{seclf}

\begin{table*} \label{lfparam}
\begin{center}
\begin{tabular}[c]{|l|c|c|c|l|}
 Band & $\phi_* (10^{-2} h^3 {\rm Mpc}^{-3})$ & $\alpha$ &
 $M_*-5{\rm log}h$ & Ref. \\
\hline
K & $1.16\pm0.10$ & $-1.09\pm0.06$ & $-23.39\pm0.05$ & 2MASS \citep{2mass} \\
$b_J$ & $1.61\pm0.08$ & $-1.21\pm0.03$ & $-19.66\pm0.07$ & 2dFGRS \citep{2df} \\
$^{0.1}g$ & $2.18\pm0.08$ & $-0.89\pm0.03$ & $-19.39\pm0.02$ & SDSS \citep{sdss} \\
$^{0.1}r$ & $1.49\pm0.04$ & $-1.05\pm0.01$ & $-20.44\pm0.01$ & SDSS \citep{sdss} \\
\end{tabular}
\caption{Schechter function parameters of the luminosity functions used.}
\end{center}
\end{table*}

\begin{figure}
\includegraphics[height=84mm,angle=270]{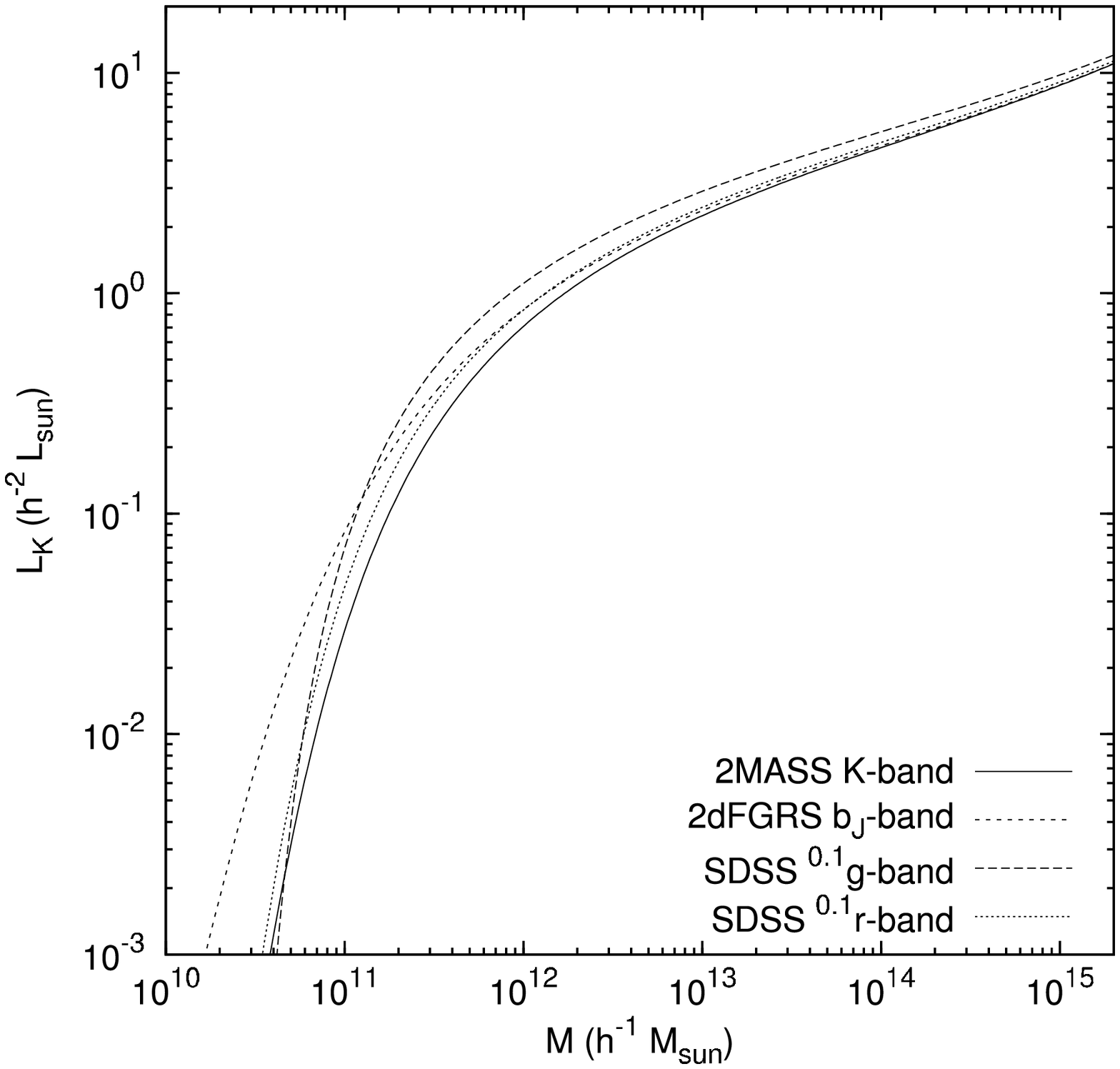}
\includegraphics[height=84mm,angle=270]{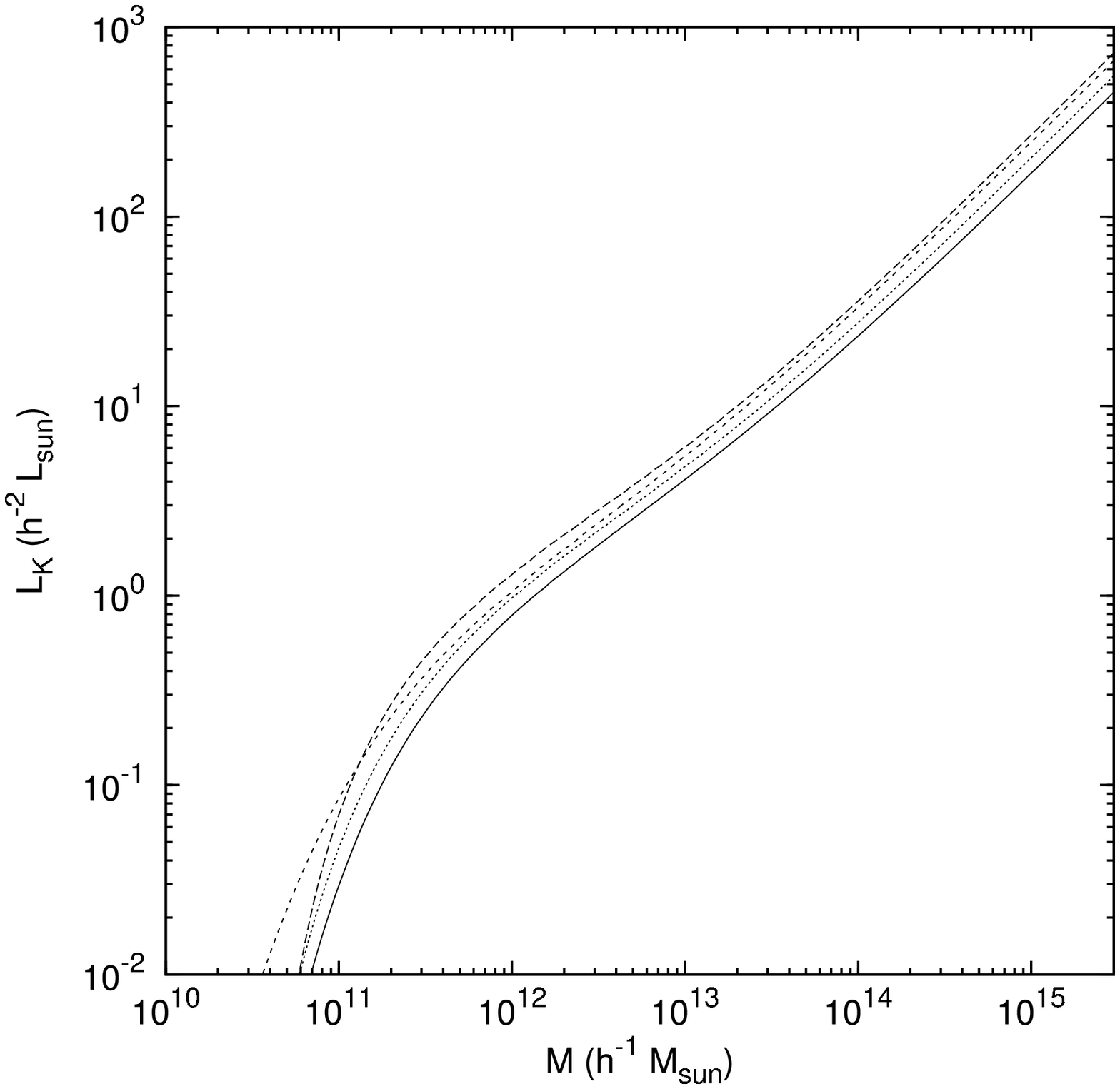}
\caption{Comparison of the galaxy (upper panel) and group (lower
panel) luminosities obtained when using the non-parametric model
described in section 4 with the different luminosity functions of table
1. The luminosities have been scaled by the characteristic luminosity
$L_*$ of the luminosity function used.}
\label{mlcompfig}
\end{figure}

In this section, we look at how our base result for the mass
luminosity relation changes depending on the luminosity function (and
consequently waveband) we use as a starting point. We use a variety
of published luminosity functions, listed in table 1, ranging from the
blue to the infrared, and repeat the analysis of the previous section
for each of these. The results for the galaxy and total luminosity are
shown in figure \ref{mlcompfig}, where the luminosity is plotted in
units of the characteristic luminosity $L_*$ of each of the luminosity
functions. 

The most striking feature of these results is that galaxy luminosity
in high mass haloes is almost independent of the luminosity function
used, with only very slight differences due to different
normalization and faint end slope. Technically, this is a consequence
of the fact that all luminosity functions considered have a bright end
cutoff of the form $exp(-L/L_*)$. In general, galaxies have an  
observed colour-magnitude relation which should be reflected in a different
mass-luminosity relation at different wavebands (as is the case here for lower
mass). In this particular case, however, almost all the galaxies for these
high mass haloes will be brightest cluster galaxies, for which an almost flat
colour-luminosity relation is expected (see e.g. \citealt{eisenstein}), 
in agreement with the model results. The scaling at this high-mass end
goes as $L\propto M^{0.28}$. This is similar to that obtained in
other studies using the conditional luminosity function formalism,
like \citet{cooray}, who find that the central galaxy luminosity
scales as $\sim M^{0.3}$ for high halo mass, while the total
luminosity of the galaxies in the halo scales as $\sim M^{0.85}$ in
the K-band; using 2dF results, \citet{yanghod} also find that the
luminosity of the central galaxy scales as $L_c\sim M^{0.25}$. A fit
to the observational K-band data gives $L\propto M^{0.26}$ \citep{lm}.

For low mass haloes and subhaloes, there are large differences between
the results obtained by using the different luminosity
functions. These are mostly a product of the characteristic
luminosity, which determines where the break in the mass luminosity
relation occurs, and the faint end slope, which determines the slope
in the relation for low mass. These results illustrate the importance
of an accurate determination of the faint end slope to obtaining a
good mass luminosity relation. Nonetheless, looking at the results for
the group luminosity it is possible to see that these differences have
only a small effect when calculating the total luminosity associated with a
high mass halo. The different results are quite close, particularly in
terms of scaling, if less so than in the individual galaxy case, and
the differences seem mostly due to the characteristic luminosity
$L_*$. There is in fact some observational evidence that the scaling of 
cluster luminosity with mass is independent of photometric band 
(see e.g., \citealt{popesso}).
Since the brightest galaxy luminosity gives a small relative
contribution for high mass systems (cf figure \ref{mlfig}), it is
possible to conclude that the main contribution to the luminosity
should come from relatively massive subhaloes with luminosities
roughly above $0.1 L_*$, below which the differences due to the low
end slope become quite significant, mostly due to different faint end
slopes in the luminosity functions. More importantly, this means that
the mass-total luminosity relation obtained by the non-parametric
model depends mostly on the more well determined of the luminosity
function parameters, $L_*$.

\subsection{Cosmological parameters} \label{seccosm}

Although the agreement between our results and the large observational
data set of brightest cluster galaxy luminosities and halo masses
shown in figure \ref{mlfig} was fairly good, it is possible to see
that our model tends to slightly underpredict the expected
luminosities. At such high masses, the subhalo contribution to the
total host number and therefore to the mass-luminosity relation is
neglegible (see figure \ref{mffig}), and, as we have shown in the
previous section, the results are also largely independent of the
luminosity function. Therefore, the relation between halo mass and
hosted brightest galaxy luminosity depends essentially on the halo
mass function, and through it on the cosmology considered. This makes
it important to look at the background cosmology used and how it
affects the results.

\begin{figure}
\includegraphics[height=84mm,angle=270]{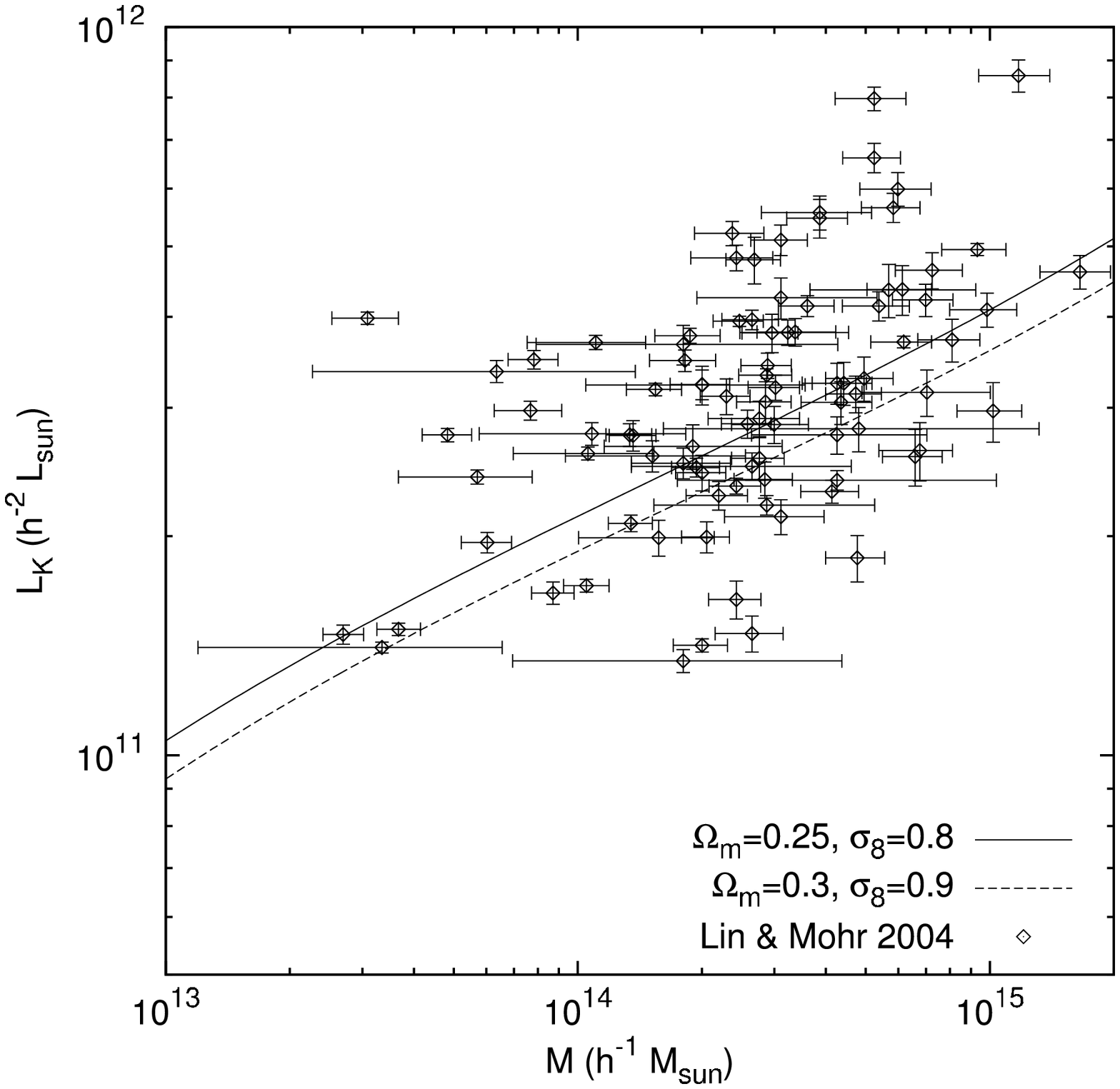}
\includegraphics[height=84mm,angle=270]{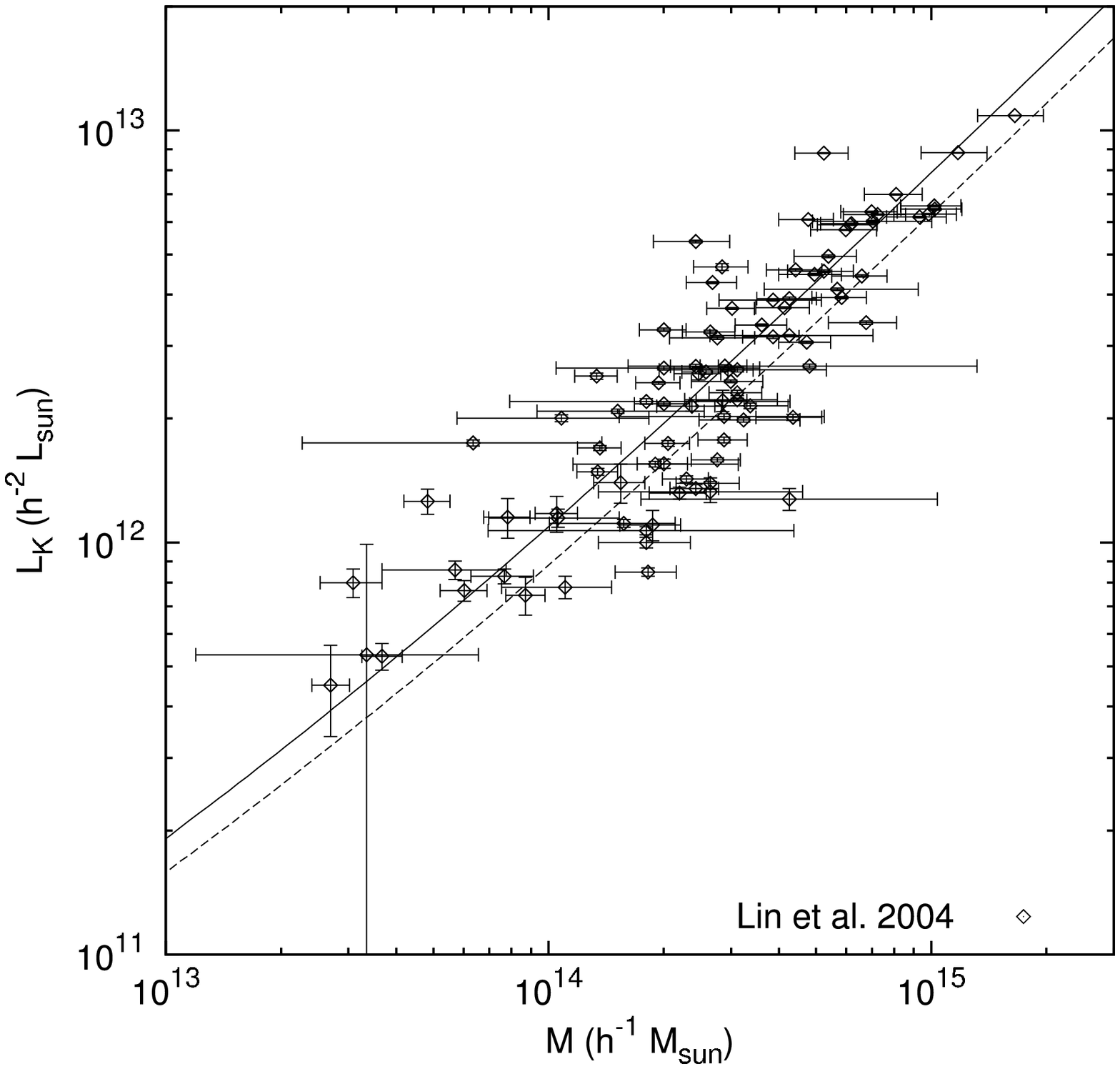}
\caption{Derived mass-luminosity relation for galaxies (upper panel)
and groups (lower panel) as a function of the cosmological model used,
for $\Omega_m=0.25$, $\sigma_8=0.8$ and $\Omega_m=0.3$,
$\sigma_8=0.9$. }
\label{mlcosm}
\end{figure}

An example of this is shown in figure \ref{mlcosm}, where we also
show the observational data for comparison. In general, it is possible
to conclude that, comparing with the observations, our model tends to
prefer lower values of $\Omega_m$ and $\sigma_8$. The technical reason
for this is quite straightfoward: lower values result in fewer high
mass haloes, which means that when comparing to observed galaxy
numbers it results in higher luminosity galaxies being associated with
the same halo mass.  For this reason, we have adopted in our
calculations a cosmological model with $\Omega_m=0.25$ and
$\sigma_8=0.8$, rather than the more standard $\Omega_m=0.3$ and
$\sigma_8=0.9$.  This combination of values lies roughly on the lower
edge of the concordance region from WMAP+SDSS measurements of
\citet{tegmark}, and near the center of that from CMB+2dF \citep{sanchez}
(see figure \ref{cosm}).

In fact, it is possible to obtain an even better agreement with the
plotted observational values if we lower $\sigma_8$ further. There are
two main reasons for not adopting such a model: first, the
\citet{tegmark} results represent the best current measured estimate
for the combination of the two; second, and more importantly, there
are further additional sources of error which are probably more
significant than what is now a very accurate determination of the
cosmological parameters. The first of these is the possibility that
the main assumptions of the model are wrong. Mainly, that is that we
cannot treat the mass luminosity relation as one-to-one and
monotonic. While this is certainly true in specific cases and we
expect there to be significant scatter about the relation, it seems a
fair assumption to make for the average relation as we are determining
here; going beyond this assumption is outside the scope of this
model. Second, and more importantly in the context of the present
paper, there is the possibility that the model is giving incorrect
results simply because the number functions we are using as a basis
for the comparison are mismatched, that is $n(m)_{\rm observed}\neq
n(m)_{\rm model}$, where the observational term applies to the data
points we are comparing our results with. Since the method is based on
counting numbers for each of these, a difference between them would
cause it to give incorrect results when comparing to observations. As
we discussed above, the cosmological model used can change the model
part of the equation. Far more likely, though, is that we are using an
incorrect estimate for the observed mass that is skewing the observed
term. We discuss this issue further in the next section.

\begin{figure}
\includegraphics[height=84mm,angle=270]{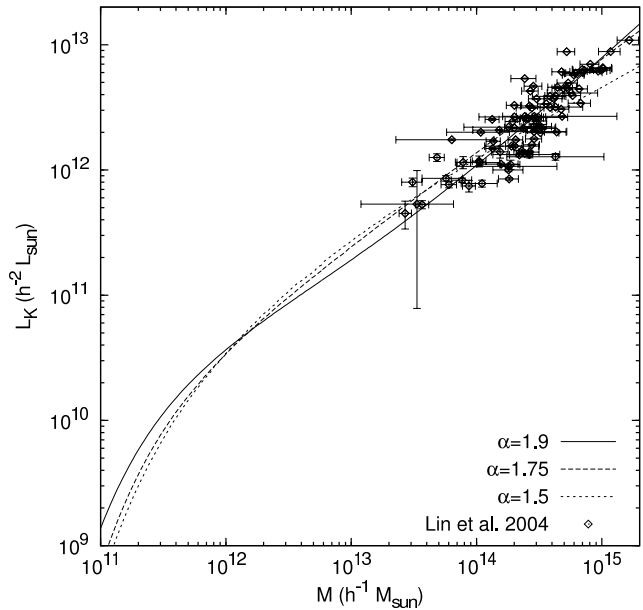}
\caption{Group luminosity obtained by varying the low mass slope of the subhalo mass function, 
equation \ref{shmf}, between $\alpha=1.5$ and $\alpha=1.9$ (our base model).} 
\label{mlshmf}
\end{figure}

It is also worthwhile noting that, once the cosmology is fixed, the
main determinant for the form of the halo mass-total luminosity
relation is the slope of the subhalo mass function. This assumes that
its normalization and cutoff are predefined, as discussed in section
3.2. Figure \ref{mlshmf} shows the result of varying this
slope. Generally, a flatter slope for the subhalo mass function
results in a flatter slope at high halo mass for the total luminosity as well. Since a
fit to the observational data gives a slope of 0.72, while we obtain
0.88, using a flatter slope for the subhalo mass function than 1.9
(quite possible in light of simulation results; see \citealt{laurie})
would likely give a better agreement with observations, depending on
possible corrections to their mass estimates, which may not be uniform.
It should be noted, however, that this in fact applies to the original 
subhalo mass function slope. How it compares with the present slope 
depends on the mass loss factor. With the one we are using in this paper, 
the slopes are mostly equal (see section 3).

\subsection{Observed mass determination} \label{secmobs}

As noted, one of the possibilities that may explain the difference
between our model results and the set of observational cluster data as
presented in figure \ref{mlfig} is a potential misestimation of the
corresponding halo mass.  First of all, there is the problem of how to
extrapolate to the virial mass.  In \citet{lms}, the authors determine
$M_{500}$, the mass enclosed in a radius where the average density is
500 the critical, from X-ray temperature measurements.  Besides the
actual observational error in the temperature, and possible errors in
the $M_{500}-T_X$ relation used (which are assumed to be accounted for
in the observational error bars), there is the problem of how to
extrapolate $M_{500}$ to the halo virial mass. Following what is done
by \citet{lms}, we assume an NFW halo with a concentration of 5 to
calculate the virial mass, except that we take the virial radius to be
at $r_{100}$ rather than $r_{200}$. A different density profile or
simply concentration would result in a different measure for
mass. Changing the concentration to a value in the range of between 4
and 7, for example, could result in as much as a 10\% shift upwards or
downwards in the estimated mass.

\begin{figure}
\includegraphics[height=84mm]{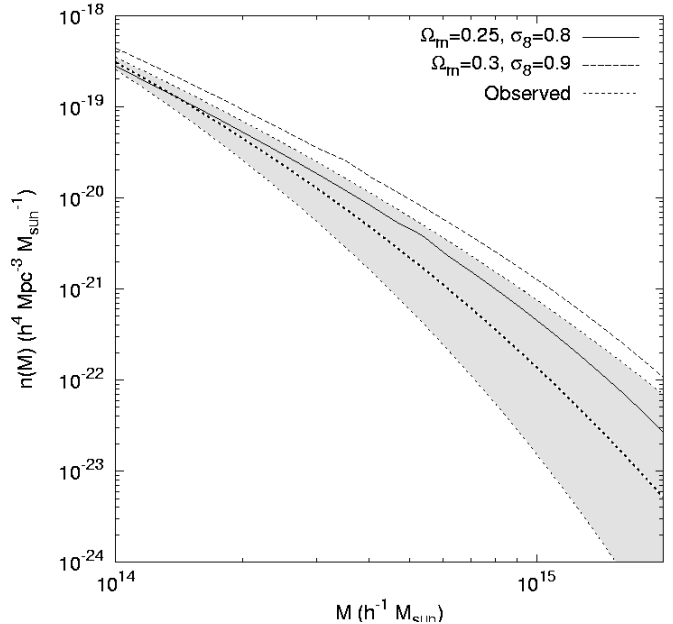}
\caption{Halo mass function for our base model, together with the
corresponding mass function for clusters as derived from the observed
luminosity function and cluster halo mass-central galaxy luminosity
relation (see text for details); the central line comes from the fit
found by \citet{lm}, the lines above and below represent the
$1\sigma$ range from the fit. Also shown is the curve for the
$\Omega_m=0.3$, $\sigma_8=0.9$ halo mass function, for comparison.}
\label{mfobs}
\end{figure}

The non-parametric model cannot work if the halo mass fucntion for the
assumed cosmology does not match observations, but there is a relatively 
straightforward way of checking how well these
mass estimates agree with the results from our model. As mentioned
previously, that is to check whether the predicted galaxy number at a
given halo mass from the observed values match the number in the model,
$n(m)_{\rm observed}=n(m)_{\rm model}$, where $n_{\rm model}(m)$ in the
high mass range we are interested in here is given by the halo mass
function. As noted, these two must match if the comparison between the model
and the data is to be meaningful, since the model is based on counting
numbers. In order to determine $n(m)_{\rm observed}$, we combine the
luminosity function with the fit to the observed halo mass-brightest
cluster galaxy luminosity relation from \citet{lm}. The resulting mass
function, together with the possible range obtained from the errors in
the fit parameters, is shown in figure \ref{mfobs}. From this we
can see that, while the halo mass function we are using does not quite
match the mass function corresponding to the observations, it is well
within the range allowed by the fit errors. If we assume that the
results from the model are essentially correct, we can then use these
curves to calculate a further correction to the estimated mass. In
fact, this corresponds to finding the transformation that matches the
fit to the observed data to our result for the average mass-luminosity
relation. 

\section{Additional checks}

In this section we calculate an additional four results which can be
derived directly from our non-parametric model, and which can be used
as an additional check on the success of the model: the mass to light
ratio, the occupation number, the group luminosity function and 
the multiplicity function. 

\subsection{Mass to light ratio} \label{secmlratio}

\begin{figure}
\includegraphics[height=84mm,angle=270]{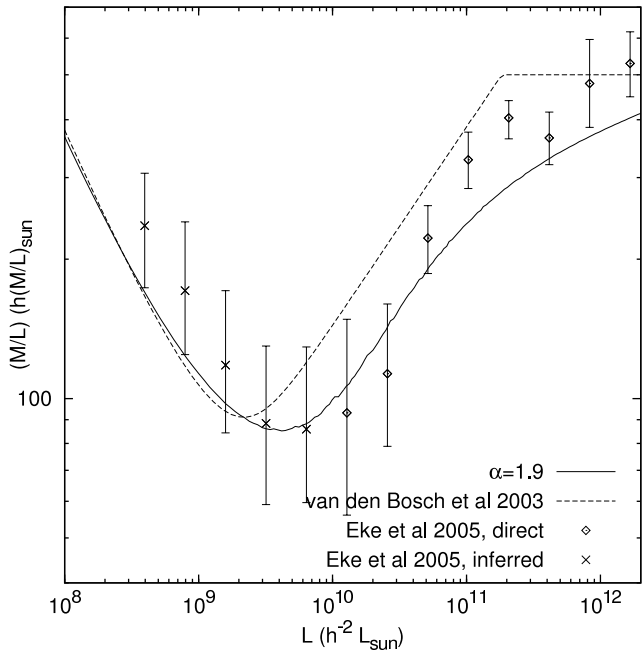}
\caption{Mass to $b_{\rm J}$-band light ratio of galaxy systems (from
isolated galaxies, to groups and clusters with growing mass). Data
points are taken from the corrected mass to light ratio calculated for
the 2PIGG catalogue of the 2dF survey by \citet{2dfglf}. These include
both the directly measured values for luminosities above $L>10^{10}
h^{-2} L_\odot$, and the values inferred by comparing the derived
group luminosity function with a theoretical mass function, for
luminosities below this threshold. The latter were extracted from the
shaded region in figure 15 of \citet{2dfglf}, where the point
represents the value of their model curve at that luminosity. The
errorbars include uncertainties related to different possible values
of $\sigma_8$.  For comparison, we also show a line for the result of
model D of \citet{BoschYangMo}.}
\label{mlratiofig}
\end{figure}

The discussed trends of the mass-luminosity relation are further
visible in figure \ref{mlratiofig}, which shows the $b_{\rm J}$-band
mass to light ratio of the entire system (that is, a group or cluster
if massive enough; this is basically a different way of plotting the
result of figure \ref{mlfig}).

As mentioned above, at the high mass end the cluster luminosity is
almost directly proportional to halo mass (roughly, $L\propto
M^{0.88}$). This means that the resulting mass to light ratio will be
almost constant, as can be seen in the figure, rising only very slowly
with halo mass, which matches well with previous results for the mass
to light ratio of clusters (e.g.,
\citealt{bcdoy,kochanek,2pigg,2dfglf}). In this case, though, the
values we obtain for the mass to light ratio seem to be slightly
smaller than the observational results at the bright end, otherwise
they seem in good agreement.  This is reflected in the values for the
cluster mass to light ratio. The value derived by \citet{fhp} is $450
\pm 100 h(M/L)_\odot$, while \citet{2pigg} obtain an average value of
$466 \pm 26 h(M/L)_\odot$; we obtain a slightly lower cluster mass to
light ratio of approximately $425 h(M/L)_\odot$ for a $10^{15} h^{-1}
{\rm M_\odot}$ halo.  Since we obtained a good agreement with the
cluster luminosity results of \citet{lms} and, as discussed above, the
mass luminosity for single galaxies is well constrained in our model,
the most likely origin for the discrepancy we find here is likely to
be the subhalo distribution. Lowering the value of the subhalo mass
function slope slightly results in a better agreement with the
observational results at both the low and high luminosity ends;
however, the agreement is poorer at intermediate luminosities, while
the slope of the derived mass to light ratio becomes steeper (cf.
results in figure \ref{mlshmf}).

We can also compare our results with the conditional luminosity
function of \citet{YangMoBosch}. Plotted in figure \ref{mlratiofig} is
the result of the average halo mass to light ratio fit with the
parameters of model D of \citet{BoschYangMo}. There is a good
agreement at the low end, while again we find some disagreement at the
high end. Note that this disagreement is most likely due to the fact
that we are adopting a different cosmology than these authors, with 
slightly lower values for $\Omega_m$ and $\sigma_8$.
 Nevertheless, the overall shape is quite similar; the small
difference is most likely due to the factor that these authors are
fitting the mass to light ratio to a double power law, with a sharp
transition to a constant at high mass, while we obtain a smooth
function.  The minimum also occurs at a slightly higher luminosity. We
are actually in better agreement with the observed values at an
intermediate range (around $10^{10}-10^{11} h^{-2} {\rm L_\odot}$),
while the opposite happens at the bright end.  between both results is
considerably better; as noted, using a flatter slope for the subhalo
mass function may help with the problem.

Figure \ref{mlratiofig} also illustrates quite well the various mass
(or, as presented in the figure, luminosity) scales which determine
galaxy properties (cf. \citealt{dekel}). There is a minimum in the mass
to light ratio at a mass scale of around $3-4 \times 10^{11} h^{-1} {\rm
M_\odot}$. This is a good match for the scale which represents a shift
in the characteristic galaxy population corresponding to a change in
the sources of gas accretion and star formation suppression (e.g.,
\citealt{dekelbirnboim}). Also, as we have already shown, the steep
slope in the mass luminosity relation means that haloes below roughly
a few times $10^9 h^{-1} {\rm M_\odot}$ will essentially be dark;
again, this matches well with the mass below which
photo-ionization is expected to suppress gas infall
\citep{br,ThoulWeinberg,dekelwoo}. At the high end, there is a noticeable
break in the mass to light ratio at a corresponding mass of around
$10^{13} h^{-1} {\rm M_\odot}$. As can be seen in figure
\ref{mlfig}, this marks the scale where we go over from
isolated galaxies with at most very small satellites to actual groups,
and once more is in good agreement with what is observed and expected
from a theoretical point of view, which predicts an upper bound for
cooling in a dynamical time at this mass scale. This is also the mass
scale at which merging is expected to become inefficient (e.g.,
\citealt{cooray}).

\subsection{Occupation number}

With the subhalo mass function given by equation (\ref{shmf}), it is
straightfoward to calculate the occupation number (that is, the number
of subhaloes in a parent halo of given mass), as a function of halo
mass $M$. This is simply given by:

\begin{equation} \label{occup}
N_s(M)=\int^\infty_{m_{min}} N(m|M) dm \, .
\end{equation}

\begin{figure}
\includegraphics[height=84mm,angle=270]{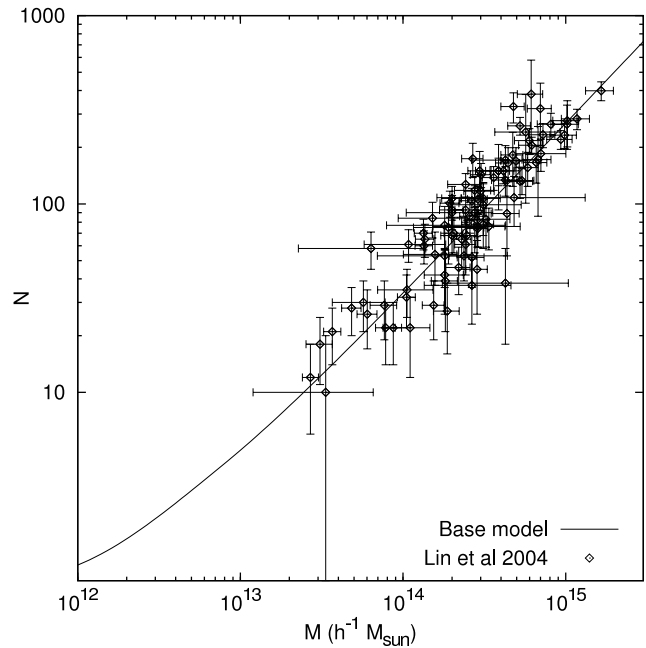}
\caption{Occupation number: average number of galaxies predicted for
a parent halo of mass $M$, given by the sum of the number of subhaloes
with galaxies above a given luminosity plus one for the central galaxy. 
The luminosity threshold adopted was $M_K<-21$, which corresponds to the
minimum used for the observed data points plotted, taken from \citet{lms}.}
\label{occupfig}
\end{figure}

\noindent The only complication is that it is necessary to specify a
minimum mass for the subhaloes, $m_{min}$; otherwise, the integral is
divergent. Since we equate subhaloes with satellite galaxies, we can
associate this minimum mass with a minimum luminosity of these
galaxies, and then $N_s(M)+1$ will give us the total number of
galaxies in a halo of mass $M$, with luminosity greater than the
minimum we are considering; this is one of the key ingredients of the
HOD models, (e.g. \citealt{bg,zz,yanghod}). These models usually take the
function in (\ref{occup}) to be the average value of the number of
satellite galaxies, with the actual number a Poisson distribution
around this average; this seems to be in good agreement with
simulation results \citep{kravtsov,zz}. Also, the occupation number of
the central galaxy is usually taken to be a step function, being 0 or
1 depending on whether the halo mass is greater than a certain
minimum; the case in our model is also similar, as we consider that a
halo hosts a galaxy with luminosity above a certain threshold if its
mass is greater than that corresponding to this luminosity from the
mass-luminosity relation. Our results for the occupation number are
shown in figure \ref{occupfig}, where we take the minimum mass to be
the necessary to host a galaxy with $M_K=-21$. The points plotted are 
taken from the cluster data of \citet{lms}, where as before we have
included a correction to the mass.

Our result is in quite good agreement with the plotted data.
At high
halo mass (approximately above $\sim 10^{13} h^{-1} M_\odot$), 
our result for the occupation number scales roughly as
$N\propto M^{0.9}$. 
Qualitatively, it also compares well with those obtained from
simulations, analytical models and observations by a variety of
authors. The first thing to note is that, unlike the result in paper
I, the occupation number is now no longer a function of $M/m_{min}$
alone: there is an extra dependence on the halo mass $M$.
Physically, this is understandable as haloes having $M/M_*>1$ are in
the process of growing and merging whereas those having $M/M_*<1$ are
decreasing in number density as they merge into larger systems. This
is a consequence of the fact that the the subhalo mass function, given in
equation (\ref{shmf}), cannot be written as a function of solely $m/M$: 
its normalization has terms which
depend only on $M$. While a direct comparison of the numbers is
complicated by the fact that they depend on what mass threshold is
being considered, the slope of the occupation number for high mass
haloes is roughly the same as that found in other studies. This matches 
well with the analytically derived
halo mass function of \citet{ogurilee}, who find a slope of 0.9 at
high $M/m\approx10^5$, and closer to 1 at lower $M/m\approx 10^2$, and
also with the results from simulations: for example, \citet{kravtsov}
find a slope very close to 1, while \citet{zz}, in fitting HOD models
to semi-analytic models of galaxy formation and smoothed particle
hydrodynamic simulations, find values for the slope between 0.97 and
1.24 (depending on the baryonic mass threshold). Results of
observational studies also seem to agree on a slope of about 1:
\citet{kochanek} obtain a relation for the number of galaxies with
$L>L_*$ scaling as $M_H^{1.1}$, while \citet{abazajian} fit a HOD
model, together with the cosmological parameters, to the projected
correlation function of a volume limited subsample of the Sloan
Digital Sky Survey (SDSS), together with CMB results, and find a good
agreement with models where the slope is fixed at a value of 1; when
they leave this as a free parameter, they obtain a result of 0.8, but
without a significant improvement in the quality of the fit.

\subsection{Group luminosity function}

Also of interest when studying clusters is the group luminosity
function, $\phi_g(L_g)$. This is the analogous of the galaxy
luminosity function, but is calculated 
for groups and clusters, and gives the number
density of these objects in a given luminosity range. We can obtain
this in our model by transforming the mass dependent halo mass
function by using the relation between halo mass and total group
luminosity:

\begin{equation} \label{grouplf}
\phi_g(L_g) dL_g= n_h(M(L_g))\frac{dM}{dL_g}dL_g \, .
\end{equation}

\noindent Here, we are implicitly assuming that a group/cluster is
made up of two components: a central galaxy hosted in the parent halo
itself, and its satellites, hosted by the subhaloes. Thus the total
luminosity is the sum of these two contributions:

\begin{equation} \label{grouplum}
L_g(M)=L(M)+\int_0^{\infty}L(m)N(m|M)dm \, .
\end{equation}

\begin{figure}
\includegraphics[height=84mm,angle=270]{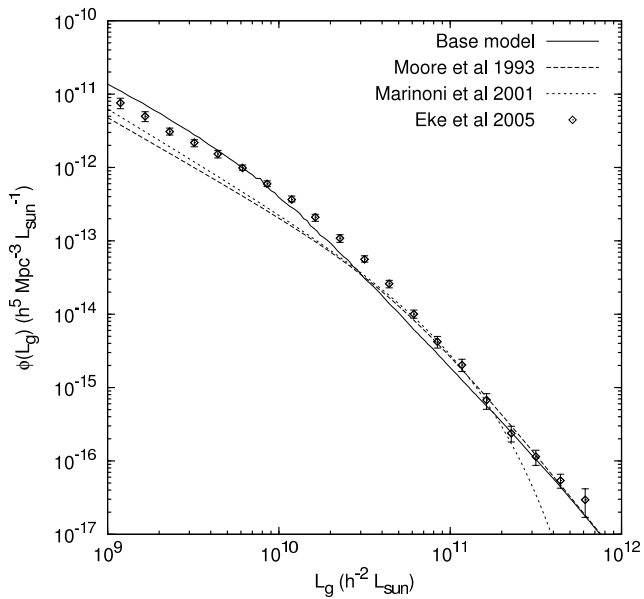}
\caption{Group luminosity function. The solid red line represents the
result  of the non-parametric model 
using equation (\ref{grouplf}); the two other lines are
observational group luminosity functions extracted from galaxy
catalogues, respectively the AGS \citep{AGS} (dotted) and the VSLF
\citep{VSLF} (dashed).}
\label{grouplumfig}
\end{figure}

\noindent This group luminosity is the same as was presented in figure
\ref{mlfig} above. Our result for the group luminosity function is
shown in figure \ref{grouplumfig}. The determination of the
observational function is usually done from galaxy catalogues by
building groups of gravitationally bound galaxies. In the figure, we
show the luminosity function fits of two different such functions: the
AGS for the CfA survey \citep{AGS} and the VSLF \citep{VSLF} group
luminosity functions. We also show data for the group luminosity
function as measured from 2dF results of \citet{2dfglf}. We have
included corrections to the AGS results by using $b_J=B_{\rm
Zwicky}-0.05$ and making the VSLF results 0.55 magnitudes fainter to
compensate for the difference between the $B$ and $b_J$ bands and
internal absorption (see comments in \citealt{2dfglf}).  We obtain a
very good agreement with the observational results. This agreement 
is markedly better than the result in paper I;
this is essentially a result of the higher group luminosity we obtain
for each halo, which in turn is caused by an increase in the subhalo
numbers at a given luminosity due to the more self-consistent way in
which we are treating mass loss.

\subsection{Multiplicity function}

\begin{figure}
\includegraphics[height=84mm,angle=270]{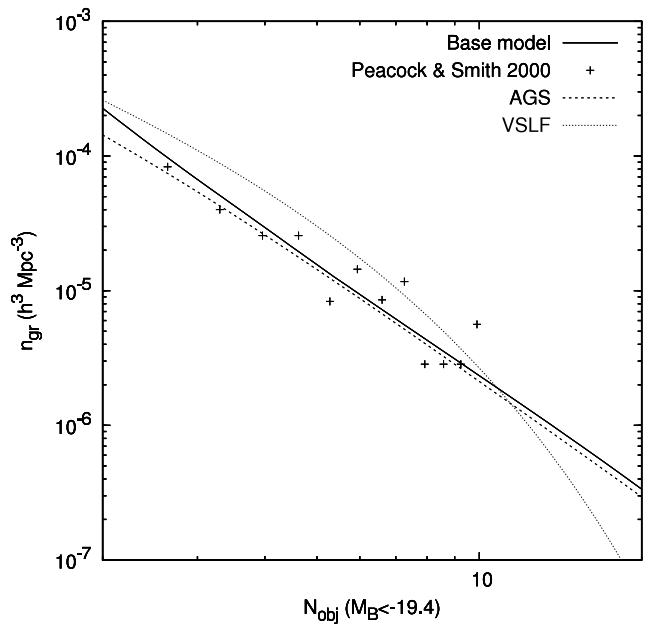}
\caption{
Multiplicity function derived from the non-parametric 
model (solid line),
based on combining the halo mass function with the total halo
occupation number, that is, the total number of galaxies, above a
given luminosity, present in a parent halo. The minimum mass for the
halo or any of its subhaloes to be counted corresponds to a magnitude
of $M_B=-19.4$. The different points are taken from the results
derived by \citet{PeacockSmith} from the CfA survey, while the two
additional lines are estimates based on the group luminosity functions shown
in figure \ref{grouplumfig}, where the magnitude limit for all of
these is also $M_B=-19.4$.}
\label{multfig}
\end{figure}

It is also possible to derive the multiplicity function, the number
density of group/clusters as a function of their richness, by a
process similar to the one described above for the group luminosity
function. To do this, we replace the group luminosity $L_g$ in
equation (\ref{grouplf}) with the number of galaxies present in the
halo. This is given by the total occupation number calculated above
and shown in figure \ref{occupfig}, that is, the number of luminous subhaloes
plus one. The only other thing that is necessary to take into account
is the limit to the galaxy luminosity we want to consider. This will
then give a lower limit to the mass of the haloes and their subhaloes
to put into the occupation number calculation. Our results for the
multiplicity function are shown in figure \ref{multfig}. We have taken
as the lower luminosity limit $M_B=-19.4$, in order to match the
observational results also shown. Due to the way in which it is built,
there is a sharp upturn at $N=1$ due to the haloes which are massive
enough to contain a galaxy but not to have subhaloes big enough to
host a galaxy themselves. This is in part a consequence of modelling
the occupation number of the parent halo as a step function, as was
done above; a more realistic model would have a smoother transtition
(see e.g. \citealt{zz} for an example of how this is done in the
context of HOD models), which would help to atenuate this.

Figure \ref{multfig} also shows some
observational data. The points were taken from the analysis of
\citet{PeacockSmith}, while the two lines were constructed from the
two group luminosity functions shown in figure \ref{grouplumfig}. To
do this, we assume that the luminosity function of the galaxies in
each group has the same shape as the general galaxy luminosity
function of the survey, a Schechter function with characteristic
luminosity $L_*$ and faint end slope $\alpha$, but with the
normalization $n$ determined by the group luminosity. By integrating
this function, the group luminosity $L_g$ can then be calculated as
$L_g=n L_* \Gamma(2-\alpha)$; this is then related to the number of
galaxies in the group above a certain threshold luminosity $L_{min}$
by

\begin{equation} \label{groupnumber}
N=\frac{L_g}{L_*}\frac{\Gamma(1-\alpha,L_{min}/L_*)}{\Gamma(2-\alpha)} \, ,
\end{equation}

\noindent where we used the expression above to replace $n$. Finally
we can use this relation between group luminosity and member numbers
in the group luminosity function to obtain the multiplicity function,
using $n(N)=\phi_g(L_g) dL_g/dN$, where $n(N)$ and $\phi_g(L_g)$ are
the multiplicity and group luminosity functions, respectively. The
galaxy numbers for both the points and the results derived from the
group luminosity functions were then multiplied by a factor of 0.66 to
take into account the difference in radius between friends of friends
estimates and the usual definition of virial radius (see
\citealt{kochanek}).

We obtain a good agreement with the observational points, better than
was the case in paper I. This is not surprising given the results we
presented above for the group luminosity function, and the root cause
is the same: the more self-consistent treatment of mass loss has
resulted in an increase in subhalo numbers at a given mass.  On the
other hand, at high member count, our result would seem to be above
the multiplicity functions calculated from the group luminosity
function. This is most likely due to the fact that we assumed above
that we can use for the luminosity function of the group galaxies the
same shape as the background luminosity function; while this should be
true for moderately dense regions (corresponding to low $N$), it is
not a very good approximation for very dense environments like clusters,
where $N$ is high (see \citealt{croton} for observational results). 
In fact, a higher $L_*$, expected in clusters with
high $N$, gives a higher value for the multiplicity function,
therefore bringing the two curves into better agreement with our
results.

\section{Conclusion}

In this paper we have presented a non-parametric model to relate the
luminosity of a galaxy to the mass of the halo or subhalo hosting
it. We start by assuming that this relation is one to one and
monotonic, and then compare their number as given by their statistical
distributions, obtained from observations and simulations,
respectively. This gives us the average relation between the
luminosity of a galaxy and the mass of the dark matter halo or subhalo
which hosts it. We can also determine the total group luminosity as a
function of halo mass by integrating over the luminosity of all the
subhaloes in a given parent.

We have argued that, to maintain the assumption of monotonicity, when
accounting for the subhaloes it is in fact necessary to consider not
their present mass, but rather their original mass just prior to
accretion into their parent halo. In order to allow for this, we
have introduced a simplified prescription for the average mass loss
factor that we can then apply to subhalo mass functions measured from
simulations to regain the original subhalo mass function. We have
noted that, however, there are two strong constraints on this
function: first, no subhalo could originally have more than half the
present mass of the parent halo, otherwise it would be, by definition,
the parent halo itself; second, it is to be expected that, since the
parent is built up by accreting and stripping mass off the subhaloes,
that the total mass in the original subhaloes equals the present mass
of the parent halo. If the mass loss factor is fairly regular (or more
precisely, its logarithmic derivative is always significantly lower than 1,
which is mostly the case with the one we presented), and assuming a
present day Schechter type subhalo mass function (which seems to be
obtained from simulation results; see e.g., \citealt{laurie}), these two
constraints mean that the only free parameter in the original subhalo
mass function would be the low mass slope, which in these
circumstances would be close to the present day one.

We find that, for high mass haloes, the mass-luminosity relation
appears to be mostly independent of the luminosity function being
used, when the luminosity is scaled to the characteristic luminosity
$L_*$. The same is true, to a rather lesser extent, of the mass-group
luminosity relation, which has an increased dependence on $L_*$. At
the low mass end, the break in the relation is associated with $L_*$,
while the slope depends significantly on the faint end slope of the
luminosity function.

Overall, our results are a good match to observations and results of
other theoretical models. We find, however, that for high mass haloes,
our results seem to slightly underestimate the luminosity of both
central galaxies and clusters when compared to the observational
results of \citet{lms} and \citet{lm}.  This small discrepancy between
our results and observational data simultaneously raises some
concern. The reason is that, as we have discussed, the mass luminosity
relation in high mass haloes is practically independent of both the
actual luminosity function used and of the subhalo population. This
means that our result depends only on the halo mass function, and
through it, on the cosmological model. We have shown that, when
comparing our results with observations, our model seems to prefer
lower values of $\Omega_m$ and $\sigma_8$; within the concordance
region of \citet{tegmark}, best results are obtained near the lower
boundary, and consequently we have used $\Omega_m=0.25$ and
$\sigma_8=0.8$ to construct our base model. While to change our result
would necessitate changing the cosmology, or more seriously, the
central assumptions underlying the model, it is far more likely that
the discrepancy is actually a product of a misestimation of the
observed halo mass.  In fact, there is some uncertainty regarding the
values of the mass cited in these studies, specifically in 
extrapolating them to the virial radius. A reasonable change
in the concentration used to make this extrapolation may account for
as much as a 10\% shift in the estimated observed mass. Further, using
the fit to the observed mass-luminosity data and the luminosity
function, we have compared the expected number of galaxies in haloes of a
given mass to the one we are using. We find that, while the agreement
is not perfect, the mass function we used is well within the error
range of the observed one. Further, assuming that our model is
correct, it is possible to view the required transformation to the
estimated mass to match the two curves as an additional correction to
it.

The situation with the total luminosity on cluster scales is
relatively similar, though here the problem might be more one of shape
rather than normalization. The slight discrepancy between our results
and the observational data may be improved if the halo mass has been
underestimated observationally or by tweaking with the cosmological
parameters. There is however an additional factor, in that the subhalo
population contributes significantly to the total luminosity. This is
represented by the original subhalo mass function, and we have argued
that its normalization and cutoff mass should be considered fixed, the
former to give the total present parent halo mass, the latter to avoid
having subhaloes which originally had more than half the parent
mass. If we assume that this function has a Schechter shape (as would
be the case with a present subhalo mass function with a Schechter
shape and a regular mass loss factor), then the only free parameter
left is the low mass slope. We have also argued, by comparing the
results obtained using different luminosity functions, that the more
important luminosity function parameter that determines the result at
this high mass end is the characteristic luminosity $L_*$, while the
faint end slope (which could be considered the most uncertain of the
parameters) does not seem to be as big a factor.  Therefore, if we
take the subhalo mass-galaxy luminosity relation (and consequently the
cosmology) to be fixed, the remaining free parameter in the model 
is the subhalo mass function low mass end slope. 

On the one hand, this means that we cannot much vary the total number
of satellite galaxies in this framework. On the other hand, our result
for the group luminosity seems to be a better match to the observed
values than that of the galaxy luminosity; if we were to
increase the latter, we might end up obtaining too high a luminosity in
groups, assuming the slope of the subhalo mass function is kept fixed.
This potential problem may however be viewed in light of the recent
results of \citet{cooraycen}, who claim that when using a similar
prescription to model the satellite galaxy luminosity function using
the subhalo mass function, it is necessary to introduce an efficiency
function lowering the number of luminous subhaloes otherwise predicted
to match the two.  There is also the possibility, which we have not
considered, that the mass-luminosity relation is different for haloes
and subhaloes; using a different relation for the latter might explain
the difference to the observational results. Since gas accretion and
mergers stop in subhaloes once they are accreted, we are led to the
conclusion that, if this were to be the case, the subhaloes would
actually be less luminous than what we are considering. Additionally,
there is the effect of the slope of the subhalo mass function. Using a
flatter slope than the 1.9 value we have used for our base model may
help with this situation, since it would flatten the slope of the
group luminosity calculated without changing the high mass galaxy
luminosity. This may be benificial, since the fit to the observed
clusters from \citet{lms} goes as $L\propto M^{0.72}$, while we get a
slightly steeper 0.88.  If we do accept the whole framework of the
model, though, this raises an intriguing possibility: the only free
parameter we have is the low mass end slope of the subhalo mass
function, and so we may be able to get a completely independent
confirmation of its value by fitting the model results to the
observational data.

Quantitavely, we find that central galaxy luminosity scales with halo
mass as $L_1\sim M^{0.28}$ for high mass haloes, fairly independently
of waveband when the luminosity is scaled by the appropriate
characteristic luminosity $L_*$, and also of the form of the subhalo
mass function. The total group luminosity scales with halo mass as
$L_{tot}\sim M^{0.88}$, also fairly independently of waveband when
appropriately scaled; a flatter subhalo mass function low mass slope results in
a flatter dependence at the high mass end. This implies that the halo
mass-to-light ratio is almost flat at high mass (and luminosity), and we
obtain a value of $425 h (M_{b_J}/L)_\odot$ for a $10^{15} h^{-1} M_\odot$
halo.  For low mass haloes, the resulting mass luminosity relation is
dependent on what waveband we are considering, but scaled luminosity
goes as $L\sim M^a$ with $a$ roughly between 4 and 4.5 for haloes or
subhaloes in the mass range $10^{9} h^{-1} M_\odot$ to $10^{10} h^{-1}
M_\odot$; for example, in the K-band used for our base model, $a\simeq
4.5$.  We also find that the occupation number, which can be derived
almost directly from the subhalo mass function, scales as $M^{0.9}$.

Finally, we should also make a note on the applicability of the
relation we obtained. First, we should stress that this is an average
relation.  As can be seen from the plotted observational data, we
expect a rather large scatter around it. While we feel that obtaining
this average relation is quite an important first step and by itself
already allows a range of applications, it is important to obtain a
model for the scatter if it is to be applied, for example, to build
mock catalogues from simulation results. In this direction, there has
already been some work on applying the base framework that we have
developed further here to a context of a conditional luminosity
function, including potential prescriptions for scatter (see work by
Cooray and collaborators, e.g. \citealt{cooraye,cooraycen}). 
Nonetheless, we feel it is relevant to have a good analysis
of the basic framework, especially since the overall simplicity of the
model makes it conceptually clear and pedagogical, while at the same
time allowing a good comprehension of the factors influencing it.

\section*{Acknowledgements}
We would like to thank George Efstathiou, Ofer Lahav and Edwin Turner 
for useful comments and suggestions.
AV acknowledges financial support from Funda\c
c\~ao para a Ci\^encia e Tecnologia (Portugal), under grant
SFRH/BD/2989/2000.


\begin{thebibliography}{99}
\bibitem[\protect\citeauthoryear{Abazajian et al.}{2005}]{abazajian}
Abazajian K. et al., 2005, ApJ, 625, 613
\bibitem[\protect\citeauthoryear{Babul \& Rees}{1992}]{br} 
Babul A., Rees M.~J., 1992, MNRAS, 255, 346 
\bibitem[\protect\citeauthoryear{Bahcall et al.}{2000}]{bcdoy}
Bahcall N.~A., Cen R., Dav{\' e} R., Ostriker J.~P., Yu Q., 2000,
ApJ, 541, 1
\bibitem[\protect\citeauthoryear{Bahcall et al.}{1999}]{triangle} 
Bahcall N.~A., Ostriker J.~P., Perlmutter S., Steinhardt P.~J., 1999, Sci, 284, 1481
\bibitem[\protect\citeauthoryear{Bailin et al.}{2005}]{bailin}
Bailin J., et al., 2005, ApJ, 627, L17
\bibitem[\protect\citeauthoryear{Bell et al.}{2003}]{bell}
Bell E.F., McIntosh D.H., Katz N., Weinberg M.D., 2003, ApJS, 149, 289
\bibitem[\protect\citeauthoryear{Benson}{2001}]{benson}
Benson A.~J., 2001, MNRAS, 325, 1039
\bibitem[\protect\citeauthoryear{Benson et al.}{2000a}]{bba}
Benson A.~J., Baugh C.~M., Cole S., Frenk C.~S., Lacey C.~G.,
2000a, MNRAS, 316, 107
\bibitem[\protect\citeauthoryear{Benson et al.}{2000b}]{bbb}
Benson A.~J., Cole S., Frenk C.~S., Baugh C.~M., Lacey C.~G.,
2000b, MNRAS, 311, 793
\bibitem[\protect\citeauthoryear{Benson et al.}{2003}]{bfb}
Benson A.~J., Frenk C.~S., Baugh C.~M., Cole S., Lacey C.~G.,
2003, MNRAS, 343, 679
\bibitem[\protect\citeauthoryear{Berlind \& Weinberg}{2002}]{bg}
Berlind A.~A., Weinberg D.~H., 2002, ApJ, 575,587
\bibitem[\protect\citeauthoryear{Berlind et al.}{2003}]{Berlindetal}
Berlind A.~A.~et al., 2003, ApJ, 593, 1
\bibitem[\protect\citeauthoryear{Bernstein \& Bhavsar}{2000}]{bb}
Bernstein J.P., Bhavsar S.P., 2000, MNRAS
\bibitem[\protect\citeauthoryear{Blanton et al.}{2003}]{sdss} 
Blanton M.~R., et al., 2003, ApJ, 592, 819 
\bibitem[\protect\citeauthoryear{Blanton et al.}{2005}]{blanton}
Blanton M.R.,  Lupton R.H., Schlegel D.J., Strauss M.A., Brinkmann J,
Fukugita M., Loveday J., 2005, ApJ, 631, 208
\bibitem[\protect\citeauthoryear{Bryan \& Norman}{1998}]{bn}
Bryan G.L., Norman M., 1998, ApJ, 495, 80
\bibitem[\protect\citeauthoryear{Bullock et al.}{2001}]{bullock} 
Bullock J.~S., Kolatt T.~S., Sigad Y., Somerville R.~S., 
Kravtsov A.~V., Klypin A.~A., Primack J.~R., Dekel A., 2001, MNRAS, 321, 559 
\bibitem[\protect\citeauthoryear{Bullock, Wechsler, \& Somerville}{2002}]{bws}
Bullock J.~S., Wechsler R.~H., Somerville R.~S., 2002, MNRAS, 329,
246
\bibitem[\protect\citeauthoryear{Cole et al.}{2000}]{cole}
Cole S., Lacey C.G., Baugh C.M., Frenk C.S., 2000, MNRAS, 319, 168
\bibitem[\protect\citeauthoryear{Colless}{1989}]{colless} 
Colless M., 1989, MNRAS, 237, 799 
\bibitem[\protect\citeauthoryear{Cooray}{2005a}]{coorayc}
Cooray A., 2005a, MNRAS, 363, 337
\bibitem[\protect\citeauthoryear{Cooray}{2005b}]{cooraye}
Cooray A., 2005b, astro-ph/0509033, submitted to MNRAS
\bibitem[\protect\citeauthoryear{Cooray \& Cen}{2005}]{cooraycen}
Cooray A., Cen R., 2005, astro-ph/0506423, submitted to ApJ
\bibitem[\protect\citeauthoryear{Cooray \& Milosavljevi\'c}{2005a}]{cooray}
Cooray A., Milosavljevi\'c M., 2005a, ApJ, 627, L85
\bibitem[\protect\citeauthoryear{Cooray \& Milosavljevi\'c}{2005b}]{coorayb}
Cooray A., Milosavljevi\'c M., 2005b, ApJ, 627, L89
\bibitem[\protect\citeauthoryear{Croton et al.}{2005}]{croton} 
Croton D.~J. et al., 2005, MNRAS, 356, 1155
\bibitem[\protect\citeauthoryear{De Lucia et al.}{2004}]{delucia}
De Lucia G., et al., 2004, MNRAS, 348, 333
\bibitem[\protect\citeauthoryear{Dekel}{2004}]{dekel}
Dekel A., 2004, astro-ph/0401503
\bibitem[\protect\citeauthoryear{Dekel \& Birnboim}{2004}]{dekelbirnboim}
Dekel A., Birnboim Y., 2004, astro-ph/0412300, submitted to ApJ
\bibitem[\protect\citeauthoryear{Dekel \& Woo}{2003}]{dekelwoo}
Dekel A., Woo J., 2003, MNRAS, 344, 1131
\bibitem[\protect\citeauthoryear{Eisenstein et al.}{2001}]{eisenstein}
Eisenstein D.J. et al., 2001, AJ, 122, 2267
\bibitem[\protect\citeauthoryear{Eke et al.}{2004}]{2pigg}
Eke V.R. et al., 2004, MNRAS, 355, 769
\bibitem[\protect\citeauthoryear{Eke et al.}{2005}]{2dfglf}
Eke V.R., Baugh C.M., Cole S., Frenk C.S., Navarro J.F., 2005, astro-ph/0510643, 
submitted to MNRAS
\bibitem[\protect\citeauthoryear{Fukugita, Hogan \& Peebles}{1998}]{fhp}
Fukugita M., Hogan C.~J., Peebles P.~J.~E., 19  98, ApJ, 503, 518
\bibitem[\protect\citeauthoryear{Gao et al.}{2004}]{gao}
Gao L., White S.D.M., Jenkins A., Stoehr F., Springel V., 2004, MNRAS, 355, 819 
\bibitem[\protect\citeauthoryear{Governato et al.}{1998}]{gbf}
Governato F., Baugh C.~M., Frenk C.~S., Cole S., Lacey C.~G.,
Quinn T., Stadel J., 1998, Natur, 392, 359
\bibitem[\protect\citeauthoryear{Hayashi et al.}{2003}]{Hayashietal}
Hayashi E., Navarro J.~F., Taylor J.~E., Stadel J., Quinn T., 2003,
ApJ, 584, 541
\bibitem[\protect\citeauthoryear{Hoekstra, Yee, \& Gladders}{2002}]{lensing} 
Hoekstra H., Yee H.~K.~C., Gladders M.~D., 2002, ApJ, 577, 595 
\bibitem[\protect\citeauthoryear{Jarrett et al.}{2000}]{2massint}
Jarrett T.H., Chester T., Cutri R., Schneider S., Skrutskie M., Huchra J.P., 2000, AJ, 119, 2498
\bibitem[\protect\citeauthoryear{Kauffmann, White \& Guiderdoni}{1993}]{kwg}
Kauffmann G., White S.D.M., Guiderdoni B., 1993, MNRAS, 264, 201
\bibitem[\protect\citeauthoryear{Kauffmann et al.}{1999a}]{kcda}
Kauffmann G., Colberg J.~M., Diaferio A., White S.~D.~M., 1999a,
MNRAS, 303, 188
\bibitem[\protect\citeauthoryear{Kauffmann et al.}{1999b}]{kcdb}
Kauffmann G., Colberg J.~M., Diaferio A., White S.~D.~M., 1999b,
MNRAS, 307, 529
\bibitem[\protect\citeauthoryear{Kravtsov et al.}{2004}]{kravtsov}
Kravtsov A.~V., Berlind A.~A., Wechsler R.~H., Klypin A.~A.,
Gottl{\"o}ber A., Allgood B., Primack J.~R., 2004, ApJ, 2004, 609, 35
\bibitem[\protect\citeauthoryear{Kauffmann, Nusser, \& Steinmetz}{1997}]{kns}
Kauffmann G., Nusser A., Steinmetz M., 1997, MNRAS, 286, 795
\bibitem[\protect\citeauthoryear{Kochanek et al.}{2001}]{2mass}
Kochanek C.~S., et al., 2001, ApJ, 560, 566 
\bibitem[\protect\citeauthoryear{Kochanek et al.}{2003}]{kochanek}
Kochanek C.~S., White M., Huchra J., Macri L., Jarrett T.~H.,
Schneider S.~E., Mader J., 2003, ApJ, 585, 161
\bibitem[\protect\citeauthoryear{Libeskind et al.}{2005}]{libeskind}
Libeskind N.I., Frenk C.S., Cole S., Helly J.C., Jenkins A., Navarro
J.F., Power C., 2005, MNRAS, 363, 146
\bibitem[\protect\citeauthoryear{Lin \& Mohr}{2004}]{lm}
Lin Y., Mohr J.J., 2004, ApJ, 610, 745
\bibitem[\protect\citeauthoryear{Lin, Mohr \& Stanford}{2004}]{lms}
Lin Y., Mohr J.J., Stanford S.A., 2004, ApJ, astro-ph/0402308
\bibitem[\protect\citeauthoryear{Magliocchetti \& Porciani}{2003}]{mp}
Magliocchetti M., Porciani C., 2003, MNRAS, 346, 186
\bibitem[\protect\citeauthoryear{Mateo}{1998}]{Mateo}
Mateo M.~L., 1998, ARA\&A, 36, 435
\bibitem[\protect\citeauthoryear{Marinoni, Hudson \& Giuricin}{2002}]{VSLF}
Marinoni C., Hudson M.~J., Giuricin G., 2002, ApJ, 569, 91
\bibitem[\protect\citeauthoryear{Meza et al.}{2003}]{meza}
Meza A., Navarro J.F., Steinmetz M., Eke V.R., 2003, ApJ, 590, 619
\bibitem[\protect\citeauthoryear{Mobasher \& Trentham}{1998}]{mt}
Mobasher B., Trentham N., 1998, MNRAS, 293, 315
\bibitem[\protect\citeauthoryear{Moore, Frenk \& White}{1993}]{AGS}
Moore B., Frenk C.~S., White S.~D.~M., 1993, MNRAS, 261, 827
\bibitem[\protect\citeauthoryear{Nagamine et al.}{2001}]{nfco}
Nagamine K., Fukugita M., Cen R., Ostriker J.~P., 2001, ApJ, 558
497
\bibitem[\protect\citeauthoryear{Navarro, Frenk, \& White}{1997}]{nfw}
Navarro J.~F., Frenk C.~S., White S.~D.~M., 1997, ApJ, 490, 493
\bibitem[\protect\citeauthoryear{Norberg et al.}{2002}]{2df}
Norberg P.~et al., 2002, MNRAS, 336, 907
\bibitem[\protect\citeauthoryear{Oguri \& Lee}{2004}]{ogurilee}
Oguri M. \& Lee J., 2004, ApJ, 355, 120
\bibitem[\protect\citeauthoryear{Peacock \& Smith}{2000}]{PeacockSmith}
Peacock J.~A.,  Smith R.~E., 2000, MNRAS, 318, 1144
\bibitem[\protect\citeauthoryear{Pearce et al.}{2001}]{pjf}
Pearce F.~R., Jenkins A., Frenk C.~S., White S.~D.~M., Thomas
P.~A., Couchman H.~M.~P., Peacock J.~A., Efstathiou G., 2001,
MNRAS, 326, 649
\bibitem[\protect\citeauthoryear{Popesso et al.}{2004}]{popesso}
Popesso P., B\"ohringer H., Brinkmann J., Voges W., York D.G., 2004, A\&A, 423, 449
\bibitem[\protect\citeauthoryear{Sanchez et al.}{2005}]{sanchez}
Sanchez A.G., Baugh C.M., Percival W.J., Padilla N.D., Cole S., Frenk C.S., Norberg P.,
2005, astro-ph/0507583, submitted to MNRAS
\bibitem[\protect\citeauthoryear{Schechter}{1976}]{schechter76}
Schechter P., 1976, ApJ, 203, 297
\bibitem[\protect\citeauthoryear{Seljak}{2000}]{seljak}
Seljak U., 2000, MNRAS, 318, 203
\bibitem[\protect\citeauthoryear{Shaw et al.}{2005}]{laurie}
Shaw L., Weller J., Ostriker J.P., Bode P., 2005, astro-ph/0509856, submitted to ApJ
\bibitem[\protect\citeauthoryear{Sheth \& Diaferio}{2001}]{sd}
Sheth R.~K., Diaferio A., 2001, MNRAS, 322, 901
\bibitem[\protect\citeauthoryear{Sheth \& Tormen}{1999}]{stmf}
Sheth R.~K., Tormen G., MNRAS, 1999, 308, 119
\bibitem[\protect\citeauthoryear{Somerville \& Primack}{1999}]{sp}
Somerville R.S., Primack J.R., 1999, MNRAS, 310, 1087
\bibitem[\protect\citeauthoryear{Somerville et al.}{2001}]{sls}
Somerville R.~S., Lemson G., Sigad Y., Dekel A., Kauffmann G.,
White S.~D.~M., 2001, MNRAS, 320, 289
\bibitem[\protect\citeauthoryear{Spergel et al.}{2003}]{wmap}
Spergel D.~N.~et al., 2003, ApJS, 148, 175
\bibitem[\protect\citeauthoryear{Tasitsiomi et al.}{2004}]{tasitsiomi}
Tasitsiomi A., Kravtsov A.V., Wechsler R.H., Primack J.R., 2004, ApJ, 614, 533
\bibitem[\protect\citeauthoryear{Tegmark et al.}{2004}]{tegmark}
Tegmark M. et al., 2004, PhysRevD, 69, 3501
\bibitem[\protect\citeauthoryear{Thoul \& Weinberg}{1996}]{ThoulWeinberg}
Thoul A.~A., Weinberg D.~H., 1996, ApJ, 465, 608
\bibitem[\protect\citeauthoryear{Trentham \& Tully}{2002}]{trentham1} 
Trentham N., Tully R.~B., 2002, MNRAS, 335, 712 
\bibitem[\protect\citeauthoryear{Trentham, Sampson \& Banerji}{2005}]{trentham2} 
Trentham N., Sampson L., Banerji M., 2005, MNRAS, 357, 783
\bibitem[\protect\citeauthoryear{Vale \& Ostriker}{2004}]{paper1} 
Vale A., Ostriker J.~P., 2004, MNRAS, 353, 189 (Paper I)
\bibitem[\protect\citeauthoryear{van den Bosch, Tormen \&
Giocoli}{2005}]{vdbshmf}
van den Bosch F.~C., Tormen G., Giocoli C., 2005, MNRAS, 359, 1029
\bibitem[\protect\citeauthoryear{van den Bosch, Yang \& Mo}{2003}]{BoschYangMo}
van den Bosch F.~C., Yang X.~H., Mo H.J., 2003, MNRAS, 340, 771
\bibitem[\protect\citeauthoryear{van den Bosch et al.}{2005}]{vymn} 
van den Bosch F.~C., Yang X., Mo H.~J., Norberg P., 2005, MNRAS, 356, 1233
\bibitem[\protect\citeauthoryear{Wechsler et al.}{2001}]{wsb}
 Wechsler R.~H., Somerville R.~S., Bullock
J.~S., Kolatt T.~S., Primack J.~R., Blumenthal G.~R., Dekel A.,
2001, ApJ, 554, 85
\bibitem[\protect\citeauthoryear{Weller et al.}{2004}]{jochen} 
Weller J., Ostriker J.~P., Bode P., Shaw L., 2004, astro-ph/0405445, submitted
to MNRAS
\bibitem[\protect\citeauthoryear{White, Hernquist, \& Springel}{2001}]{whs}
White M., Hernquist L., Springel V., 2001, ApJ, 550, L129
\bibitem[\protect\citeauthoryear{Yang et al.}{2003}]{yangwl}
Yang X.H., Mo H.J., Kauffmann G., Chu Y.Q., 2003, MNRAS, 339, 387
\bibitem[\protect\citeauthoryear{Yang, Mo \& van den Bosch}{2003}]{YangMoBosch}
Yang X.~H., Mo H.~J., van den Bosch F.~C., 2003, MNRAS, 339, 1057
\bibitem[\protect\citeauthoryear{Yang et al.}{2005}]{yanghod}
Yang X., Mo H.J., Jing Y.P., van den Bosch F.C., 2005, MNRAS, 358, 217
\bibitem[\protect\citeauthoryear{Yoshikawa et al.}{2001}]{ytj}
Yoshikawa K., Taruya A., Jing Y.~P., Suto Y., 2001, ApJ, 558, 520
\bibitem[\protect\citeauthoryear{Zehavi et al.}{2005}]{zehavi}
Zehavi I. et al., 2005, ApJ, 630, 1
\bibitem[\protect\citeauthoryear{Zentner et al.}{2005}]{zentner}
Zentner A.R., Berlind A.A., Bullock J.S., Kravtsov A.V., Wechsler R.H., 2005, 
astro-ph/0411586, submitted to ApJ
\bibitem[\protect\citeauthoryear{Zheng et al.}{2002}]{zt}
Zheng Z., Tinker J.~L., Weinberg D.~H., Berlind A.~A., 2002, ApJ,
575, 617
\bibitem[\protect\citeauthoryear{Zheng et al.}{2004}]{zz}
Zheng Z. et al., 2004, astro-ph/0408564 
\end{thebibliography}
\end{document}